\def\aj{AJ}%
\def\araa{ARA\&A}%
\def\apj{ApJ}%
\def\aap{A\&A}%
\def\aapr{A\&A~Rev.}%
\def\mnras{MNRAS}%
\def\nat{Nature}%
\def\physrep{Phys.\ Rep.}%
\documentclass[aps,onecolumn,prd,showpacs,showkeys,preprintnumbers,superscriptaddress,nobibnotes,floatfix,longbibliography,notitlepage,nofootinbib]{revtex4-1}

\pdfoutput=1
\usepackage{amsmath}
\usepackage{amsfonts}
\usepackage{amssymb}
\usepackage{mathrsfs}
\usepackage{graphicx}
\usepackage{xcolor}
\usepackage{xspace}
\usepackage{placeins}
\usepackage{fontawesome}

\usepackage{array}
\usepackage{bigints}
\usepackage{tabularx}

\usepackage[colorlinks]{hyperref}
\hypersetup{
    colorlinks,
    linkcolor={red!50!black},
    citecolor={blue!50!black},
    urlcolor={blue!80!black}
}

\usepackage{hyperref}
\usepackage{multirow}
\usepackage{upgreek}
\usepackage[capitalise]{cleveref}

\newcommand{\eq}{Eq.~}
\newcommand{\eqs}{Eqs.~}

\newcommand{\fig}{Fig.~}
\newcommand{\Fig}{Fig.~}

\newcommand{\photInjRate}{\frac{dn_{\gamma}}{dEdt}}
\newcommand{\kDiss}{k_{\mathrm{H}_2}}
\newcommand{\kDetach}{k_{\mathrm{H}^-}}
\newcommand{\sigv}{\langle\sigma v \rangle}
\newcommand{\fdm}{f_{dm}}
\newcommand{\pann}{p_\textrm{ann}}
\newcommand{\pmax}{p_\textrm{ann,max}}

\newcommand{\cobColour}{blue\xspace}

\newcommand{\reionColour}{yellow\xspace}

\newcommand{\pradaColour}{light red\xspace}

\newcommand{\angelColour}{dark red\xspace}

\newcommand{\haloRegionColour}{light blue\xspace}

\begin{document}

\title{Supermassive black hole seeds from sub-keV dark matter}

\author{Avi Friedlander}
\email{avi.friedlander@queensu.ca}
\affiliation{Department of Physics, Engineering Physics and Astronomy, Queen's University, Kingston ON K7L 3N6, Canada}
\affiliation{Arthur B. McDonald Canadian Astroparticle Physics Research Institute, Kingston ON K7L 3N6, Canada}

\author{Sarah Schon}
\email{sqs7027@psu.edu}
\affiliation{Institute for Gravitation and the Cosmos, The Pennsylvania State University, University Park, PA 16802, USA}
\affiliation{Department of Physics, The Pennsylvania State University, University Park, PA, 16802, USA}

\author{Aaron C. Vincent}
\email{aaron.vincent@queensu.ca}
\affiliation{Department of Physics, Engineering Physics and Astronomy, Queen's University, Kingston ON K7L 3N6, Canada}
\affiliation{Arthur B. McDonald Canadian Astroparticle Physics Research Institute, Kingston ON K7L 3N6, Canada}
\affiliation{Perimeter Institute for Theoretical Physics, Waterloo ON N2L 2Y5, Canada}

\begin{abstract}
Quasars observed at redshifts $z\sim 6-7.5$ are powered by supermassive black holes which are too large to have grown from early stellar remnants without efficient super-Eddington accretion. A  proposal for alleviating this tension is for dust and metal-free gas clouds to have undergone a process of direct collapse, producing black hole seeds of mass $M_\textrm{seed}\sim10^5 M_\odot$ around redshift $z\sim17$. For direct collapse  to occur, a large flux of UV photons must exist to photodissociate molecular hydrogen, allowing the gas to cool slowly and avoid fragmentation. We investigate the possibility of sub-keV mass dark matter decaying or annihilating to produce the UV flux needed to cause direct collapse. To do so, we calculate the produced UV flux from dark matter annihilations and decays within the gas cloud's halo and compare these to the requirements of the UV spectrum found by previous hydrodynamical simulations. We find that annihilating dark matter with a mass in the range of $13.6 \textrm{ eV} \le m_{dm} \le 20 \textrm{ eV}$ can produce the required flux while avoiding existing constraints. A non-thermally produced dark matter particle which comprises the entire dark matter abundance requires a thermally averaged cross section of $\sigv \sim 10^{-35}$ cm$^3/$s. Alternatively, the flux could originate from a thermal relic which comprises only a fraction $\sim10^{-9}$ of the total dark matter  density. Decaying dark matter models which are unconstrained by independent astrophysical observations are unable to sufficiently suppress molecular hydrogen, except in gas clouds embedded in dark matter halos which are larger, cuspier, or more concentrated than current simulations predict. Lastly, we explore how our results could change with the inclusion of full three-dimensional effects. Notably, we demonstrate that if the $\mathrm{H}_2$ self-shielding is less than the conservative estimate used in this work, the range of both annihilating and decaying dark matter models which can cause direct collapse is significantly increased.
\end{abstract}

\maketitle

\section{Introduction} \label{sec:Introductino}
Whilst far from complete, a compelling narrative of galactic cosmology has emerged over the last century \cite{1998A&A...331L...1S, Benson_2010, 2017ARA&A..55...59N, 2018PhR...780....1D, 2020NatRP...2...42V}. Quantum-scale fluctuations, seeded very early in the Universe's density field, are believed to have slowly grown until becoming massive enough to collapse, forming the first halos. These gravitational wells then continued to accumulate matter through accretion of surrounding material as well as through mergers with other halos. Once sufficient gas is retained within a halo, star formation becomes possible provided the gas is able to cool on brief enough timescales. Due to the absence of metals and dust in the primordial gas there are believed to be notable differences in the manner in which the very first, Population III (Pop III), stars formed as compared to their more contemporary counterparts. We refer the reader to \cite{2004ARA&A..42...79B, 2013ASSL..396..103G, 2013RPPh...76k2901B, 2016PhR...645....1B} for reviews. 

Illuminating the precise details of this very early epoch of star and galaxy formation is challenging, as direct observational signatures are beyond the reach of current surveys. Nonetheless, the highest redshift observations to date have provided some reference points to help frame the emerging narrative. Some of the oldest and most curious objects documented to date include optically bright quasars observed at $z\sim 6$-$7.5$ \cite{2011Natur.474..616M, 2015Natur.518..512W}. Quasars are the ultra luminous active galactic nuclei (AGN) of galaxies, powered by material being accreted onto a central  supermassive black hole (SMBH). This arrangement is not in of itself at odds with the current expectation of how galaxies should look, but rather surprising due to how early such massive objects seem to have assembled. The black holes at the core of some of the oldest quasars have inferred masses in the range $10^8$--$10^{10}$ $M_\odot$ \cite{2015Natur.518..512W, 2001AJ....122.2833F}. Even in the unlikely case of sustained near-Eddington accretion rates, it is challenging to grow a SMBH from a conventional stellar black hole seed by redshift 7 or within the first billion years after the Big Bang \cite{2001ApJ...552..459H, 2017PASA...34...31V}. 

To explain how the evidently present, high redshift quasars came into existence, one can invoke either an increased growth rate, greater black hole seed mass or some combination of the two \cite{2010A&ARv..18..279V, 2011arXiv1105.4902N, 2016PASA...33...51L, 2020ARA&A..58...27I}. Pop III stellar black hole seeds are expected to have masses around $100$ $M_\odot$. Given that only 275 quasars have been observed at redshift $z>6$ and only 8 have been observed at $z>7$~\cite{2022arXiv221206907F}, it is not unfathomable that sustained, near-Eddington accretion rates (or alternatively intermittent super-Eddington rates) are possible in these rare cases. However, feedback could potentially deplete gas reservoirs and self-limit accretion in $\sim 10^{6}$ $M_\odot$ halos which are expected to host the first stars. This effect may be less disruptive in more massive halos though in turn these objects are increasingly rare with increasing mass and redshift. In dense stellar cluster environments, accelerated mass acquisition via mergers may also be possible.  In this scenario there are questions to what degree gravitational kicks disperse black holes and stars and limit the rate of these mergers \cite{2004ApJ...613...36H}.
    
Recent cosmological simulations including such feedback disfavour increased accretion rates as an explanation for the observed quasars, instead indicating that it is easier to explain their size by beginning with more massive black hole seeds of around $10^4$--$10^6 \;M_\odot$ \cite{Zhu2020}. One massive seed candidate emerges from the direct collapse black holes (DCBH) scenario, which proposes formation in the same high redshift halos as the first stars and galaxies \cite{2003ApJ...596...34B, 2014MNRAS.440.1263Y}. Unlike the case of Pop III star formation, the gas contracts quasi-isothermally towards the halo's center at high enough temperatures as to prevent fragmentation into smaller clumps, instead forming a $\sim 10^5$ $M_\odot$ black hole directly via gravitational instability or via a massive star. In order for DCBHs to form it is therefore imperative that the cooling rates are low enough to not allow the gas to reach the Jeans mass locally. Primordial gas, being devoid of dust and metals, relies on atomic and molecular hydrogen cooling channels. The latter may be suppressed by a strong enough background of ultraviolet (UV) or near-infrared photons with energies in the range $0.76 \textrm{ eV} \le E \le 13.6$ $\mathrm{eV}$. As DCBHs are expected to form very early during the cosmic dawn, the presence of a nearby galaxy is invoked to provide the necessary number of dissociating photons. Extensive work has been done to quantify the UV flux required to form DCBHs and what conditions this places on the luminosity and distance of nearby star forming galaxies \cite{Haiman:1996rc,Omukai:2000ic,Shang_2010,Wolcott-Green2011,Sugimura:2014sqa,Glover:2015osa,Regan2016,agarwal2016new,wolcott2017beyond,Luo_2020}. However, the presence of a nearby galaxy raises the potential question of contamination via metals and dust which would prevent DCBHs from forming. Alternatively, turbulence driven direct collapse \cite{Latif2022}, self-interacting dark matter~\cite{Pollack:2014rja,Feng:2020kxv,Xiao:2021ftk}, primordial black holes~\cite{Khlopov:2004sc,Kawasaki:2012kn,Kohri:2014lza,PhysRevD.92.023524,Bernal:2017nec,Ashoorioon:2022raz}, superconducting cosmic strings \cite{Brandenberger:2021zvn,Cyr:2022urs}, gas streaming velocities~\cite{2017MNRAS.471.4878S,Hirano:2017wbu}, and dark stars \cite{2016RPPh...79f6902F} have also been invoked as possible mechanisms to enhance the formation of DCBHs or SMBH seeds (see \S 6.2 \cite{2020ARA&A..58...27I} for further discussion).

In this work we propose sub-keV mass dark matter with a suitable coupling to the Standard Model as a source of UV photons to facilitate the formation of DCBHs. Models of particle dark matter generically require some portal to the Standard Model, and photons are predicted as byproducts of dark matter self-annihilation or decay to Standard Model final states in nearly every realization. Dark matter is ubiquitous to the halos in which SMBHs are expected to form, allowing photons to be injected directly into the gas clouds. The impact of dark matter energy injection in early gas clouds has been studied but primarily has a focus on heavier dark matter with a mass $> 1\textrm{ GeV}$ which produces photons with energies much above the UV range required to produce DCBHs \cite{2009ApJ...705.1031S, Schon:2014xoa,2010MNRAS.406.2605R}. By having the UV photons sourced within the gas cloud's halo, complications relating to requiring a nearby star forming region and the potential dust and metal contamination that can be associated with those nearby stars can be avoided. Another advantageous feature of dark matter producing the UV photons is that the effect of molecular hydrogen self-shielding is potentially reduced, in turn lowering the required UV flux. In this work we investigate the requirements on dark matter particle models and the host halos for DCBHs to form. We additionally explore the differences that exist in this setup compared to the traditional scenario of an externally-sourced UV flux.

This article is structured as follows: in Section \ref{sec:CritCurve} we outline the relevant gas chemistry and critical conditions required for direct collapse, Section \ref{sec:PhotonSpec} covers the treatment of energy injection from dark matter, followed by Sections \ref{sec:halo} and \ref{sec:particleModels} which respectively discuss our dark matter halo and dark matter particle models. Section \ref{sec:results} shows constraints and results, and we conclude with further discussion and conclusions in Section \ref{sec:discussion} and Section \ref{sec:conclusion}.

\section{Gas cloud cooling due to dark matter}
\subsection{The Direct Collapse Critical Curve} \label{sec:CritCurve}
A necessary condition for DCBHs to form is that the early gas cloud cools slowly enough for the gas to collapse in a uniform way, without fragmenting into smaller objects. Molecular hydrogen cools the gas rapidly, and therefore the overall abundance of $\mathrm{H}_2$ must be suppressed, leaving atomic cooling as the primary cooling channel. By studying the impact that injected photons have on the gas cloud chemistry, the condition restricting $\mathrm{H}_2$ abundance can be mapped onto a condition on the minimum flux of injected photons.

At densities below $10^3$ cm$^{-3}$, the dominant molecular hydrogen formation mechanism is the two step chemical process \cite{Stancil1998,Lepp_2002JPhB} 
\begin{align} 
\mathrm{H} + e^- &\to \mathrm{H}^- + \gamma ,\\ 
\mathrm{H} + \mathrm{H}^- &\to \mathrm{H}_2 + e^- .
\end{align}
This leads to two direct $\mathrm{H}_2$ suppression mechanisms. These are \textit{photodissociation}:
\begin{equation}
    \mathrm{H}_2 + \gamma \to \mathrm{H} + \mathrm{H},
\end{equation}
and \textit{photodetachment}:
\begin{equation}
    \mathrm{H}^- + \gamma \to \mathrm{H} + e^-.
\end{equation}
Photodissociation of $\mathrm{H}_2$ requires photons in the Lyman-Werner (LW) range of $11.2 \textrm{ eV} \le E \le 13.6 \textrm{ eV}$ whereas photodetachment of $\mathrm{H}^-$ can occur from photons with energy as low as $0.76$~eV~\cite{shapiro1987hydrogen}. The suppression of $\mathrm{H}_2$ thus depends on both the intensity and spectral shape of the the UV flux.

Early studies of direct collapse mapped the $\mathrm{H}_2$ suppression requirement onto a critical flux intensity in the LW energy band which must be exceeded for DCBHs to form~\cite{Haiman:1996rc,Omukai:2000ic,Shang_2010,Wolcott-Green2011,Sugimura:2014sqa,Glover:2015osa}. However more recent work has refined this to include the spectral dependence by moving from a critical intensity to a critical curve in the space of photodissociation~($\kDiss$) and photodetachment~($\kDetach$) rates per target molecule~\cite{agarwal2016new,wolcott2017beyond,Luo_2020}. The combination of $\kDiss$ and $\kDetach$ parameterise all relevant information about the UV photon flux below $13.6$ eV. 

The photodetachment rate per $\mathrm{H}^-$ ion can be calculated by convolving the photodetachment cross section, $\sigma_{\mathrm{H}^-}(E)$, with the photon flux, $J(E)$, such that
\begin{equation} \label{eq:kdetach}
\kDetach = \int_{0.76\textrm{eV}}^{13.6\textrm{eV}} \sigma_{\mathrm{H}^-}(E) J(E) dE.
\end{equation}
The photodetachment cross section is approximated by \cite{shapiro1987hydrogen} 
\begin{equation}
    \sigma_{\mathrm{H}^-}(E) \approx 4.31\times10^{-18} \frac{\tilde{E}^3}{(0.0555 + \tilde{E}^2)^3} \textrm{ cm}^2,
\end{equation}
with
\begin{equation}
    \tilde{E}=\sqrt{\frac{E - 0.754 \textrm{ eV}}{13.6\textrm{ eV}}} .
\end{equation}

Due to the large number of molecular excitation modes, the photodissociation cross section in the LW energy range is not smooth. Rather than calculating the convolution of flux and cross section as in Eq. \eqref{eq:kdetach}, the photodissociation rate per $\mathrm{H}_2$ molecule is typically approximated by assuming a flat spectrum. This results in a rate of \cite{wolcott2017beyond}
\begin{equation} \label{eq:kDissApprox}
    \kDiss \approx 1.39\times10^{-12} \text{ s}^{-1} \bigg(\frac{J_{LW}}{10^{-21}\textrm{erg s}^{-1}\textrm{Hz}^{-1}\textrm{cm}^{-2}\textrm{sr}^{-1}} \bigg),
\end{equation}
where $J_{LW}$ is the average flux in the LW energy range. Both $\kDetach$ as described in \eq \eqref{eq:kdetach} and $\kDiss$ as described in \eq \eqref{eq:kDissApprox} have a linear dependence on the UV flux intensity. Therefore, increasing the intensity without changing the spectral shape will increase both photodissociation and photodetachment rates, resulting in a suppression of $\mathrm{H}_2$.

While \eq \eqref{eq:kDissApprox} is important for parameterising the UV photon flux, it only describes the photodissociation rate in the optically-thin regime. As the gas cloud collapses, existing molecular hydrogen within the cloud will reduce the range of LW photons propagating through the cloud. We relate the physical photodissociation rate  to the optically-thin rate, $\kDiss(N_{\mathrm{H}_2}=0,T)$, via~\cite{Wolcott-Green2011}
\begin{equation}
    \kDiss(N_{\mathrm{H}_2}, T) = \kDiss(N_{\mathrm{H}_2}=0,T) f_{sh}(N_{\mathrm{H}_2}, T),
\end{equation}
where $\kDiss(N_{\mathrm{H}_2}=0,T)$ is determined using \eq \eqref{eq:kDissApprox}.
The shielding fraction, $f_{sh}$ depends on the column density of hydrogen molecules $N_{\mathrm{H}_2}$ and the cloud temperature $T$ \cite{Wolcott-Green2011}:
\begin{equation}
    f_{sh}(N_{\mathrm{H}_2}, T) = \frac{0.965}{(1+x/b_5)^{1.1}} + \frac{0.035}{(1+x)^{0.5}}\exp[-8.5\times10^{-4}(1+x)^{0.5}],
\end{equation}
where $x\equiv N_{\mathrm{H}_2}\times10^{14}$ cm$^{-2}$, $b_5 \equiv b/10^5$ cm s$^{-1}$, and $b$ is the Doppler broadening parameter.

Calculating the $\mathrm{H}_2$ column density is computationally expensive so it is typically approximated based on the local number density and some characteristic length, $L_c$~\cite{Schaye:2001hu}:
\begin{equation} \label{eq:ColumnDensity}
    N_{\mathrm{H}_2} = n_{\mathrm{H}_2} L_c.
\end{equation}

\begin{figure}[hbt]
    \centering
    \includegraphics[width=\textwidth]{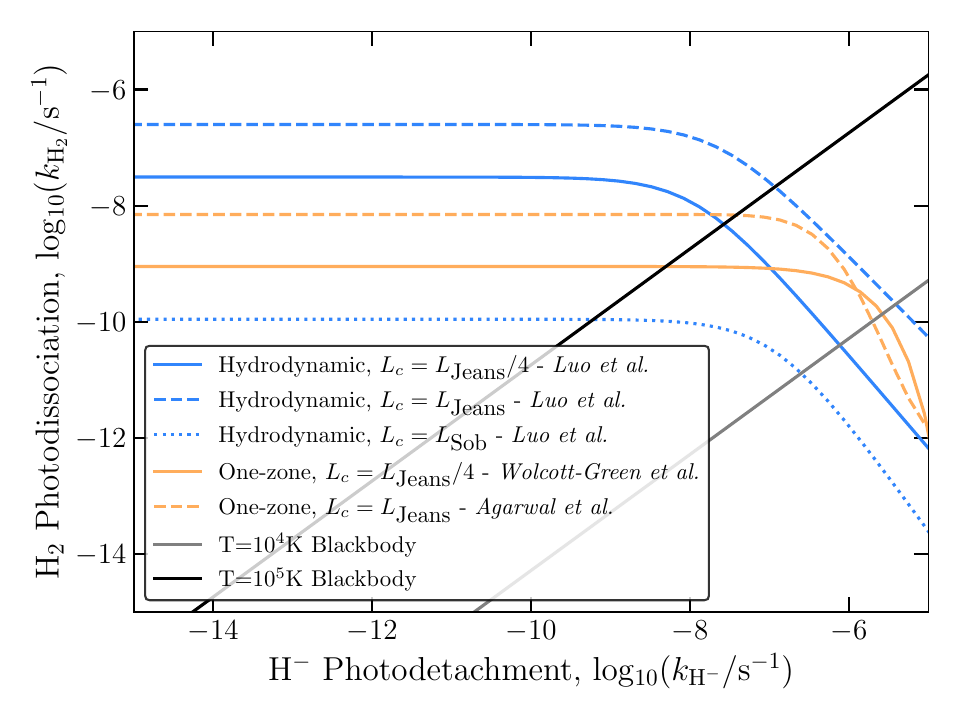}
    \caption{Direct collapse critical curves from previous papers. Photodetachment and photodissociation rates above the critical curve results in $\mathrm{H}_2$ suppression whereas reaction rates below the critical curve allow $\mathrm{H}_2$ to cool the gas efficiently. Line colour indicates the simulation method and line style indicates the $\mathrm{H}_2$ self-shielding approximation used to find the critical curve. Orange critical curves were found using a one-zone model which tracks gas density and temperature in the centre of the cloud \cite{wolcott2017beyond,agarwal2016new}. The \haloRegionColour curves were found using a 3D hydrodynamic simulation \cite{Luo_2020}. The dashed lines use the Jeans length as the self-shielding characteristic length in \eq \ref{eq:ColumnDensity} whereas the solid curves use one quarter of the Jeans length and the dotted curve uses a Sobolev-like length. The gray and black lines indicate the photodetachment and photoionization rates that would be produced by a blackbody photon spectrum with temperature of $10^4$ K and $10^5$ K respectively.} 
    \label{fig:CritCurves}
\end{figure}

As shown in \fig \ref{fig:CritCurves}, multiple works have produced critical curves in the space of $\kDetach$ and $\kDiss$ above which DCBHs can form.
The main differences between these curves are: simulation method for the gas collapse, $\mathrm{H}_2$ self-shielding approximation for calculating $L_c$ in \eq \eqref{eq:ColumnDensity}, and small changes in the chemical network. The first two critical curves, shown as orange in \Fig \ref{fig:CritCurves}, used a one-zone model where the entire gas cloud is parameterised solely by a single density and temperature~\cite{agarwal2016new, wolcott2017beyond}. \textit{Agarwal et al.} \cite{agarwal2016new} treat $\mathrm{H}_2$ self-shielding by using the Jeans length, $L_\textrm{Jeans}$, as the characteristic length in \eq \eqref{eq:ColumnDensity}. The Jeans length describes the size of a gas cloud in hydrostatic equilibrium and is given by $L_\textrm{Jeans} = c_s \sqrt{\pi/(G\rho)}$ where $c_s$ is the speed of sound, $G$ is the gravitational constant, and $\rho$ is the gas density. Therefore, using $L_c = L_\textrm{Jeans}$ in \eq \eqref{eq:ColumnDensity} is a good approximation for gas clouds close to hydrostatic equilibrium \cite{Schaye:2001hu}. However, the Jeans length has been found to overestimate the column density in 3D simulations \cite{Wolcott-Green2011} leading \textit{Welcott-Green et al.} \cite{wolcott2017beyond} to use $L_\textrm{Jeans} / 4$ as the characteristic length. 

More recently, \textit{Luo et al.} \cite{Luo_2020} updated the calculation of the direct collapse critical curve via a 3D hydrodynamic simulation of the gas collapse. They obtained three critical curves, shown in \haloRegionColour in \fig \ref{fig:CritCurves}, using three different approximations of the self-shielding characteristic length: the Jeans length, one quarter of the Jeans length, and a Sobolev-like length. The Sobolev-like length is given by $L_\textrm{Sob}=\rho/|\nabla\rho|$ and was inspired by previous studies which showed it may be a good approximation in regions with $\mathrm{H}_2$ densities below $10^4$ cm$^{-3}$ \cite{Gnedin:2008fc,Wolcott-Green2011}. They compared the accuracy of \eq \eqref{eq:ColumnDensity} using each approximation and found the reduced Jeans length ($L_\textrm{Jeans}/4$) to most closely replicate the correct column density. Therefore, in this work we mainly use the \textit{Luo et al.} critical curve from a hydrodynamic simulation using the reduced Jeans length self-shielding approximation as a threshold condition (the \haloRegionColour solid curve in \fig \ref{fig:CritCurves}) to determine whether a sufficient UV flux is produced to cause a DCBH to form. When the injected photons originate from the dark matter halo rather than from outside the gas cloud, the column density will likely be reduced. We will thus also present results using the Sobolev-like self-shielding approximation (the \haloRegionColour dotted curve in \fig \ref{fig:CritCurves}). \textit{Luo et al.} found that for most of the inner 10~pc of the gas cloud, where self-shielding is most important, using the Sobolev-like length yields a column density that is an $\mathcal{O}(1)$ factor smaller than the reduced Jeans length~\cite{Luo_2020}. Therefore, our results which use the $L_\mathrm{Sob}$ self-shielding critical curve serve as a demonstration of the impact that reduced self-shielding could have.

%---------------------------------

\subsection{Energy Injection from Dark Matter} \label{sec:PhotonSpec}

In order for direct collapse to occur, a flux of LW photons must be injected into the collapsing gas cloud. For the calculations of the photodissociation rate in \eq~\eqref{eq:kDissApprox} to apply, the injected photon flux must be continuous and relatively flat in the LW range from $11.2$ to $13.6$ eV, similar to a $10^4-10^5$~K blackbody spectrum. This can be achieved through a dark matter (sub)component with a sub-keV mass which either decays or annihilates to an extended spectrum of photons, i.e. a 3- or more body final state.

Whereas most studies of direct collapse black hole formation focus on an external UV photon source, if dark matter produces the UV photon spectrum then the photons are produced directly inside the collapsing gas cloud. This change of the photon source location could potentially affect the direct collapse critical curves in the previous section. The photon flux would be less uniform when coming from the dark matter halo, potentially allowing for rapid molecular cooling in the outer regions of the gas cloud even when $\mathrm{H}_2$ formation is suppressed in the central region. Additionally, $\mathrm{H}_2$ self-shielding would likely be reduced, especially in the dense central region because photons would be absorbed \textit{in situ}. These differences and their expected impact are discussed in more detail in Section \ref{sec:critCurveValid}.

The photon flux at a point in the halo can be found by integrating the injection rate at all other locations in the halo. The flux observed at location $\vec{r}$ in the optically thin regime is thus
\begin{equation} \label{eq:spatialFlux}
    J(\vec{r},E) = \frac{E}{4\pi}\int dV' \photInjRate(\vec{r'}, E) \frac{1}{(\vec{r'}-\vec{r})^2} .
\end{equation}
where $dn_\gamma/dEdt$ is the differential injection rate of photons from dark matter decay or annihilation per unit volume. This flux has a spatial dependence that would not be present in traditional direct collapse cases therefore for comparison to the critical curves, a location must be chosen. We calculate the flux at the centre of the halo which, if the halo and gas cloud centres are aligned, corresponds to the densest region of the gas cloud where the black hole seed would be expected to form.  \eq \eqref{eq:spatialFlux} then simplifies to
\begin{equation} \label{eq:centreFlux}
    J(E) = E \int dr' \photInjRate(r',E).
\end{equation}
Since the integrand in \eqref{eq:centreFlux} depends on both the dark matter halo and microphysics model, we must now turn to the details of this modelling.

%------------------------------------

\subsection{Dark Matter Halos} \label{sec:halo}
Energy injection from dark matter depends on the abundance and distribution of dark matter throughout the gas cloud. However, there are no observations of gas clouds or their dark matter halos at early redshifts of $z \geq 10$. We therefore rely on cosmological simulations to estimate the expected shape and size of these halos. In this section we discuss how we parameterise the dark matter distribution and what we expect the shape of the halos to be based on previous cosmological simulations.

To avoid the central divergence, we parameterise the dark matter distribution with an Einasto profile~\cite{einasto1965construction}
\begin{equation} \label{eq:EinastoProfile}
    \rho_{dm}(r) = \rho_0 \exp\Bigg(\frac{-2}{\alpha}\bigg[\bigg(\frac{r}{r_0}\bigg)^\alpha -1 \bigg]\Bigg).
\end{equation}
At a given redshift and halo size, the density and scale parameters $\rho_0$ and $r_0$ are related to the halo mass $M_{200}$, the concentration parameter $c$, and a free shape parameter $\alpha$. The halo mass is defined as the total mass contained within the radius $r_{200}$ where $r_{200}$ extends to the point where the dark matter density is 200 times the critical density. The concentration parameter is normally defined via the Navarro-Frenk-White (NFW) \cite{Navarro:1996gj} profile
\begin{equation} \label{eq:NFWProfile}
\rho_{NFW} = \frac{4 \rho_s}{\frac{r}{r_s}\bigg(1+\frac{r}{r_s}\bigg)^2},
\end{equation}
and relates the NFW radius and density parameters to the physical $M_{200}$ and $r_{200}$ by
\begin{equation} \label{eq:NFWrConc}
 r_s = r_{200}/c
\end{equation}
and
\begin{equation}\label{eq:NFWrhoConc}
\rho_s =\frac{  c^3(1+c) M_{200}  }{ 16 \pi r_{200}^3 [-c + (1+c) \ln(1+c)] }.
\end{equation}
To map these quantities to our Einasto parameterisation, we use \eq \eqref{eq:NFWrConc} and \eq \eqref{eq:NFWrhoConc} to determine the NFW parameters and then obtain the Einasto parameters $\rho_0$ and $r_0$ via a fit.\footnote{This step of determining NFW parameters and fitting to an Einasto profile could be skipped by using the Einasto concentration parameter, $c_\textrm{Ein}=r_{200}/r_0$, as in e.g. Ref. \cite{Angel:2015ilq}. However, we use the NFW concentration parameter and keep this step to allow for comparison with  mass-concentration relations found in earlier work, which only produced results for NFW halos.} 

As a reference point for the average Einasto $\alpha$ parameter, we use the relationship found by {\it Gao et al.} \cite{Gao:2007gh}:
\begin{equation}
    \alpha(M_{200},z) = 0.0095 \nu(M_{200},z)^2 + 0.155 .
\end{equation}
{\it Dutton and Maccio} \cite{Dutton:2014xda} provide a fit for the dimensionless peak height, $\nu$:
\begin{equation}
    \log_{10}\bigg(\nu(M_{200},z=0)\bigg) = -0.11 + 0.146 M_\textrm{log12} + 0.0138  M_\textrm{log12}^2 + 0.00123M_\textrm{log12}^3,
\end{equation}
where 
\begin{equation} \label{eq:Mlog12}
    M_\textrm{log12}\equiv\log_{10}\bigg(\frac{M_{200}}{10^{12} M_\odot h^{-1}}\bigg),
\end{equation}
 and 
\begin{equation}
    \frac{\nu(M_{200},z)}{\nu(M_{200},z=0)}=0.033 + 0.79(1+z) + 0.176\exp(-1.356z).
\end{equation}

We will obtain results using two different mass-concentration relations from the literature to show the potential impact of halo concentration on DCBH formation. The details of these relations can be found in Appendix \ref{sec:cMeqs}. For small halos, simulations agree that the average halo concentration decreases as a halo's mass and redshift increase. {\it Prada et al.} found that for especially large halos or at large redshift, concentration starts to increase \cite{Prada2012}. Therefore in our study of large halos at high redshift, we use {\it Prada et al.} as a reference point for highly concentrated but realistic dark matter halos. Alternatively, the $c-M$ relationship from {\it Angel et al.} did not find that same uptick in halo concentration \cite{Angel:2015ilq}. We use their fit as a more conservative benchmark. The mass-concentration parameter relation is in general fit to results from cosmological simulations, with complimentary observations at low redshifts coming from strong and weak gravitational lensing, X-rays, and galaxy kinematics in clusters \cite{Bhattacharya2013}. Extending these fits towards low halos masses is limited by resolution of simulations and direct observations being predominantly suitable for large mass halos. Both these limitations are exacerbated for high redshift halos in addition to the more frequent kinematic disruption of early halos. However, numerical fits may be compared to semi-analytic models \cite{2015Correa} which derive the mass-concentration relation by evoking halo formation theory and can be readily extended to lower masses and high redshifts. Therefore, the use of the fits described above is well motivated.

Universal dark matter halo profiles used in this work generally assume a general degree of spherical symmetry as well as relaxation into a fully virialised state, which is broadly the case for the mature halos observed in the late Universe. In contrast there is growing evidence that high redshift halo display a far more disrupted range of morphologies \cite{Poole2016, Sasaki2014, Druschke2018} both in their dark and gas components. Physical properties of these early halos such as non-spherical morphology and inhomogeneities pre-seeded in the gas via clumping of the dark matter during the halo collapse, as well as dynamical factors such as turbulence, non-uniform gas accretion and angular momentum are not captured by our simplified one-zone model but could nonetheless play a role in determining whether direct collapse of the gas is possible. Turbulence created by non symmetrical collapse or cold accretion can act as an additional source of pressure support and therefore enhance the chances of direct collapse black hole formation \cite{Latif2022nat}. Similarly, the broader environment the halo is situated in may provide favourable environments for DCBH formations \cite{Schauer2017} via increased streaming velocities. On the other hand any kind of asymmetry could further gas fragmentation, leading to the formation of stars rather than a single massive object. Along similar lines, clumpiness of the gas can also seed fragmentation \cite{Bromm2002}  while any kind of dark matter substructure may provide a boost in annihilation power and therefore and more efficient molecular hydrogen dissociation. In turn, non-uniform distribution of the gas will also lead to variation in the chemical evolution as denser region will more efficiently self-shield from external radiation, while more readily absorbing the photons created within. We look to explore these factors and how they interrelate with each other in later work, using a full three dimensional hydro model. In this work, we confine ourselves to modeling the density within the central most part of the halo in which the simplified assumptions of our one-zone model hold and allow us to gain a general understanding at what mass scales this DCBH formation mechanism would be effective.

\begin{figure} [htb]
    \centering
    \includegraphics[width=\textwidth]{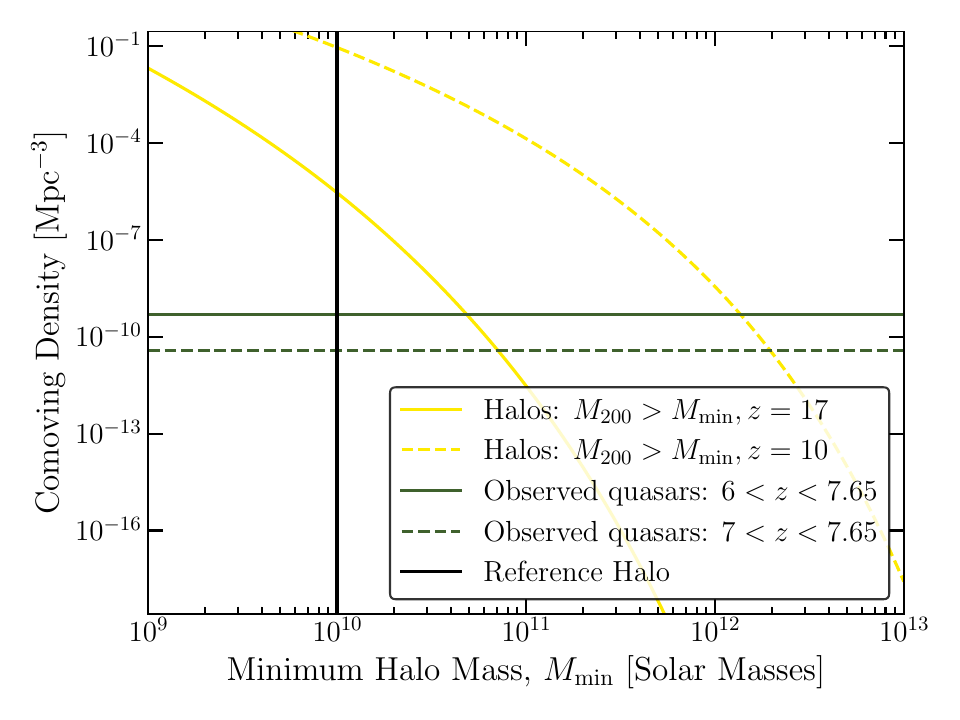}
    \caption{Comparison of density of massive halos at early redshifts to the density of observed quasars per comoving volume. The yellow curves show the integrated halo mass function above a minimum mass $M_\mathrm{min}$ representing the maximum number of halos larger than a given mass that could have collapsed by redshift $z=10$ (dashed) or $z=17$ (solid). The green horizontal lines depict the average density of observed quasars in the redshift ranges $6<z<7.65$ (solid) and $7<z<7.65$ (dashed). The black vertical line shows the reference halo mass used throughout this work.}
    \label{fig:QuasarDensity}
\end{figure}

Larger halos produce a larger flux but are rarer. If a gas cloud must collapse by redshift $z_\mathrm{col}$ to form a high-redshift SMBH, the abundance of early quasars can be related to the halo mass function at $z_\mathrm{col}$ by
\begin{equation} \label{eq:quasarDensity}
    n_\mathrm{quasars}(z) = f_q \int^\infty_{M_\mathrm{min}}dM_{200} \frac{dn_\mathrm{halo}}{dM_{200}}(M_{200}, z_\mathrm{col})\,, 
\end{equation}
where $n_\mathrm{quasars}$ is the density of quasars at redshift $z$ per comoving volume, $M_\mathrm{min}$ is the minimum mass of a halo which can collapse to form a seed large enough to become a quasar, and $dn_\mathrm{halo}/dM_{200}$ is the density of halos of mass $M_{200}$ per comoving volume. The prefactor, $f_q$, is the fraction of halos above mass $M_\mathrm{min}$ which result in quasars taking into account requirements on the initial halo such as a sufficiently large concentration and the absence of dust, the fraction of seeds which grow to become SMBHs, and the fraction of SMBHs which have sufficiently large luminosities to be an observable quasar. In general it is expected that $f_q \ll 1$. We set a conservative limit for the largest value $M_\mathrm{min}$ could be by ensuring that with $f_q \sim 1$, $n_\mathrm{quasars}$ found with \eq~\eqref{eq:quasarDensity} is larger than the observed density of high-redshift quasars. The 275 quasars observed in the range $6\le z\le 7.65$ corresponds to an average density per comoving volume of $5\times 10^{-10}$ Mpc$^{-3}$ and the 8 quasars observed in the range $7\le z\le 7.65$ corresponds to $4\times10^{-11}$ Mpc$^{-3}$. 

Figure \ref{fig:QuasarDensity} compares the density of massive halos at redshifts $10$ and $17$ respectively as the yellow dashed and solid curves to the density of observed quasars shown as green horizontal lines. Appendix \ref{app:hmf} describes the fitting functions used to determine the halo density. Throughout this work we conservatively use $z=17$ for a reference redshift for which we study DCBH formation. This allows us to consistently compare with the the {\it Luo et al.}~\cite{Luo_2020} critical curves in \fig~\ref{fig:CritCurves} which were found using halos at $z \sim 17$. Additionally, black hole seeds of $M_\mathrm{seed} \gtrsim 10^5 M_\odot$ formed at $z\sim 17$ are able to grow sufficiently to account for the observed high-$z$ SMBHs~\cite{Zhu2020}. As seen in Figure \ref{fig:QuasarDensity} large halos become much more abundant by later redshifts such as $z=10$. If future cosmological simulations demonstrate that black hole seeds formed at redshift $z=10$ are able to grow to become the observed high redshift quasars then the production mechanism presented in this work will be viable for a wider range of dark matter models than we present.

We will use a reference mass of $M_\textrm{ref} = 10^{10} M_\odot$. At $z=17$ the density of halos with $M_{200} \geq 10^{10}M_\odot$ is much larger than the observed density of quasars. This allows for $f_q \ll 1$ and for our conclusions to remain robust as future telscopes are expected to observe additional quasars. However, in the case where these reference halos are unlikely to produce a sufficient UV flux to cause direct collapse to occur, we will explore how large halos would have to be in order to produce DCBHs. Figure \ref{fig:QuasarDensity} demonstrates that so long as $M_{min} < 5\times10^{10} M_\odot$, the observed quasar abundance can be reproduced with $f_q < 1$.

\begin{figure} [htb]
    \centering
    \includegraphics[width=0.45\textwidth]{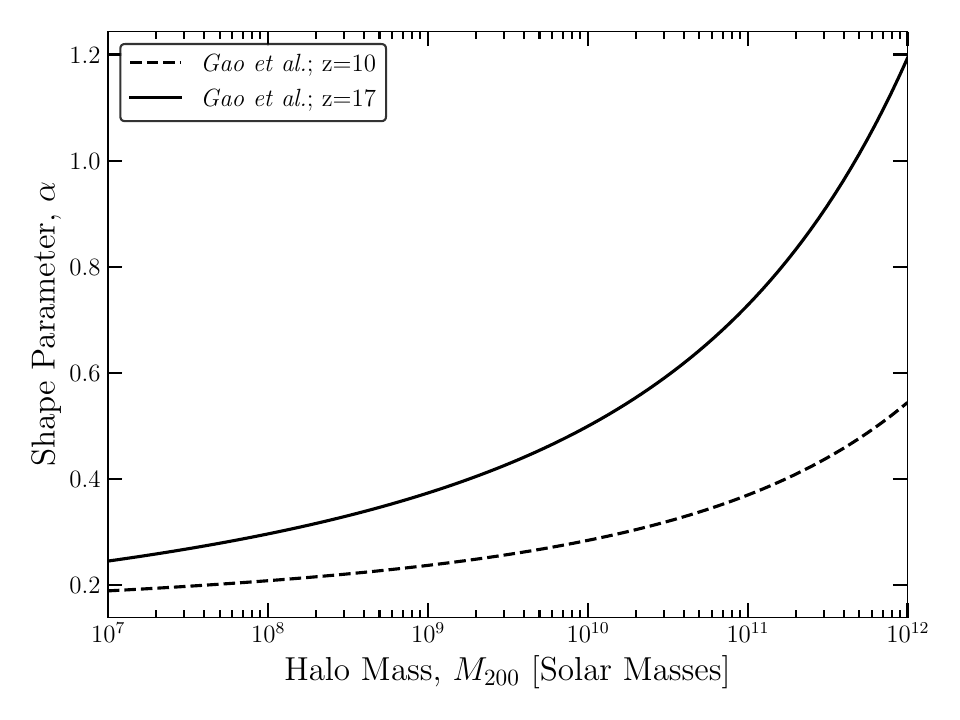}
    \includegraphics[width=0.45\textwidth]{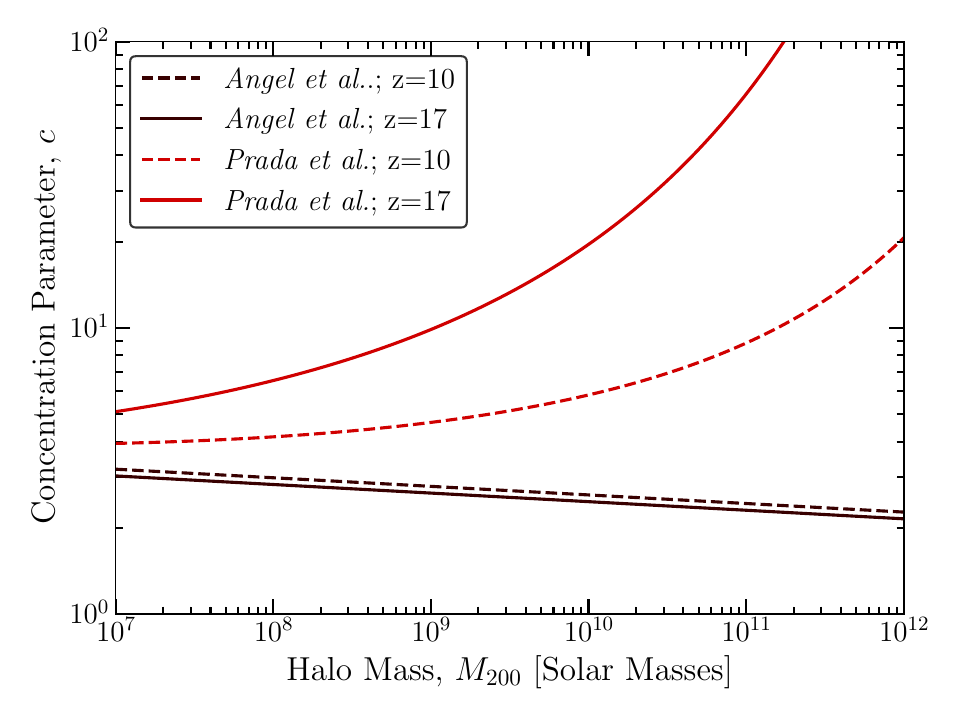}
    \caption{Left: Average Einasto shape parameter, $\alpha$, from  {\it Gao et al.} \cite{Dutton:2014xda}. Right: Halo concentration fits from {\it Prada et al.} \cite{Prada2012} (\pradaColour) and {\it Angel et al.} \cite{Angel:2015ilq} (\angelColour). Solid curves in both panels show relationships at $z=17$ and dashed curves show relationships at a later redshift of $z=10$.}
    \label{fig:HaloParam}
\end{figure}

Figure \ref{fig:HaloParam} shows the $c-M$ and $\alpha-M$ relationships for average halos at redshifts $z=10$ and $z=17$. More massive halos and halos at higher redshift tend to have larger $\alpha$ values, meaning that they are more cored, whereas smaller halos tend to be cuspier. Halos at the reference mass of $10^{10} M_\odot$ have an average shape of $\alpha = 0.5$ at $z=17$ but that decreases to $\alpha = 0.28$ at $z=10$. The $c-M$ relationships differ substantially depending on which fit is used. At a halo of mass $10^{10} M_\odot$, the {\it Prada et al.} fit predicts an average concentration of $c=20$ but the {\it Angel et al.} fit predicts a much lower concentration of $c=2.5$. This difference in halo concentrations has a large impact on the resulting flux. 

%------------------------------------

\subsection{Photon production from dark matter} \label{sec:particleModels}
We now turn to production mechanisms of LW photons from dark matter. We will examine two broad classes of dark matter energy injection models: decay and annihilation.

\subsubsection{Decay}
Dark matter must be stable on cosmological time scales, and stringent limits exist on its lifetime based on the effects of electromagnetic energy injection from decay at various cosmological epochs, though these bounds can be significantly relaxed if the decaying species is a subcomponent. Many models, including light dark matter models such as sterile neutrinos and axion-like particles predict some amount of decay. For a dark matter component with mass $m_{dm}$ and lifetime $\tau$, the in-situ photon injection rate is given by
\begin{equation}
    \photInjRate(E,r) = \frac{f_\gamma \fdm \rho_{dm}(r)}{\tau m_{dm} } \frac{dN}{dE}(E)
\end{equation}
where $f_\gamma$ is the fraction of energy released as photons in each decay, $\fdm$ is the fraction of dark matter comprising this decaying component, $\rho_{dm}(r)$ is the dark matter halo energy density, and ${dN}/{dE}$ is the photon spectrum per decay. The flux in the centre of the halo in \eq \eqref{eq:centreFlux} then becomes an integral over the halo profile:
\begin{equation} \label{eq:decayFlux}
    J(E)  = \frac{f_\gamma \fdm}{\tau m_{dm} } E\frac{dN}{dE}\int dr' \rho_{dm}(r') .
\end{equation}
Using an Einasto profile \eqref{eq:EinastoProfile}, the integral over dark matter density can be performed analytically:
\begin{equation}
    J(E) = \frac{f_\gamma \fdm}{\tau m_{dm} } E\frac{dN}{dE}\rho_0 r_0 g_{dec}(\alpha),
\end{equation}
where 
\begin{equation}
g_{dec}(\alpha)\equiv \bigg(\frac{e^2\alpha}{2}\bigg)^{1/\alpha} \Gamma\bigg(1+\frac{1}{\alpha}\bigg),
\end{equation}
and $\Gamma(x)$ is the \textit{gamma} function.

The microphysical particle details contribute to the prefactor in \eq \eqref{eq:decayFlux} such that
\begin{equation}
    J \propto \frac{f_\gamma \fdm}{\tau m_{dm} } E\frac{dN}{dE}.
\end{equation}
The parameters $f_\gamma$ and $\fdm$, are dependent on a specific dark matter model's decay process and mechanism for dark matter production in the early Universe. We use reference parameters for decaying dark matter such that $f_\gamma = \fdm = 1$. Both the UV flux and constraints on the lifetime, $\tau$, have a linear dependence on $f_\gamma$ and $\fdm$ so changes in either parameter would shift the constraint on $\tau$ and region of interest for DCBHs by the same amount. This means that as $f_{\gamma}$ or $\fdm$ change, the relevant range of lifetimes would change but qualitative conclusions regarding the viability of decaying dark matter producing DCBHs would remain the same.

The spectral shape, ${dN}/{dE}$, is dependent on the phase space distribution of the decay products. A monoenergetic line is produced in the case of a two-body final state. As we have noted, however, dissociation of $\mathrm{H}_2$ requires an extended spectrum across the LW band. This is because the cross section consists of discrete lines corresponding to the energy required to transition to different excited states. Unless the monoenergetic spectrum from a two-body dark matter decay happened to overlap with one of the transition energies, the decay products would not contribute to photodissociation. The exact shape of a $(n > 2)$-body decay will depend on the number of outgoing particles, the particle couplings, and the masses of any invisible final states. We will use two phenomenological parameterisations for the outgoing photon spectrum: a ``parabola'', and a ``box'' spectrum. The parabola is modeled as:
\begin{equation} \label{eq:parabolaSpectrum}
    E\frac{dN}{dE}\bigg|_\textrm{parabola} = 6\bigg[\frac{E}{m_{dm}} - \bigg(\frac{E}{m_{dm}}\bigg)^2\bigg] \Theta\bigg(1-\frac{E}{m_{dm}}\bigg),
\end{equation}
where $\Theta(x)$ is the Heaviside step function. The box model is:
\begin{equation} \label{eq:boxSpectrum}
    E\frac{dN}{dE}\bigg|_\textrm{box} =  \Theta\bigg(1-\frac{E}{m_{dm}}\bigg).
\end{equation}
The parabola spectrum would be the exact shape for a decay to three photons where the scattering amplitude is energy independent. While the box spectrum does not map as neatly onto realistic fundamental models, it provides an intuitive comparison to the parabola model which is qualitatively distinct. This allows us to demonstrate that the results are robust to some significant changes in the energy spectrum of decay products.

\begin{figure}[hbt]
    \centering
    \includegraphics{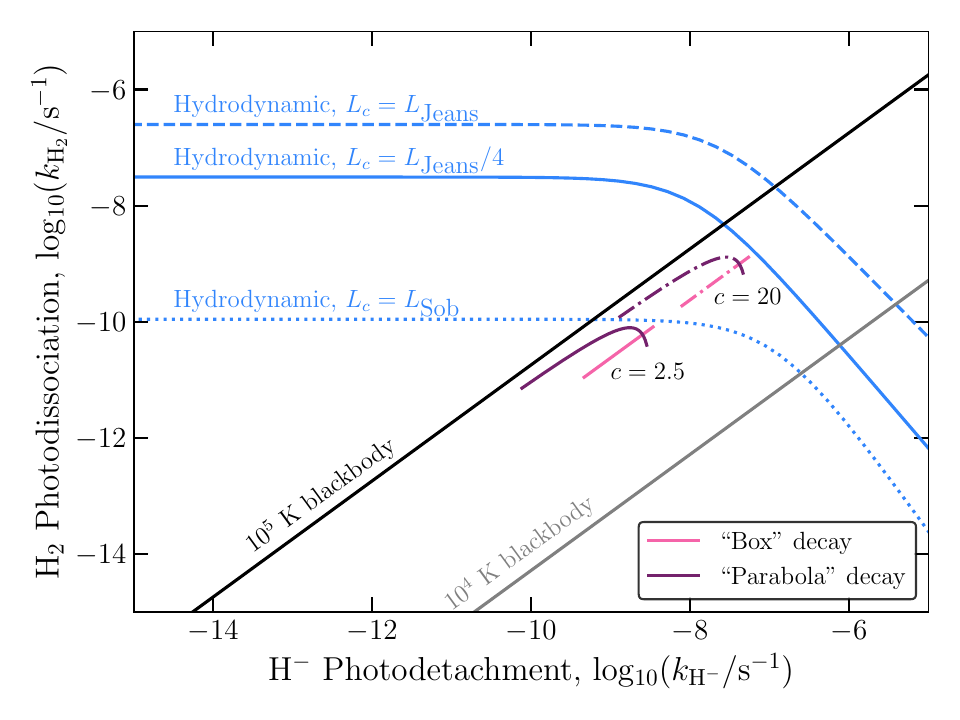}
    \caption{Comparison of the $\mathrm{H}^-$ photodetachment rate ($\kDetach$) and $\mathrm{H}_2$ photodissociation rate ($\kDiss$) produced by dark matter decay to a parabola spectrum of photons described in \eq \eqref{eq:parabolaSpectrum} versus a model which decays to a box spectrum \eqref{eq:boxSpectrum}. The \haloRegionColour curves as well as the black and gray blackbody lines are the same as in \fig \ref{fig:CritCurves}. The pink and purples curves show the reaction rates for a photon spectrum produced from a $10^{10} M_\odot$ mass halo, $\alpha=0.5$ and dark matter particle with a lifetime of $\tau=3\times10^{23}$ s. The solid curves show a halo with a concentration parameter of $c=2.5$ whereas the dot-dashed curves show a more concentrated halo with $c=20$. The length of the lines shows how the reaction rates change with dark matter mass, $m_{dm}$, in the range of $13.6 \textrm{ eV} \le m_{dm} \le 100 \textrm{ eV}$. The end of the pink and purple lines with larger reaction rates corresponds to smaller masses with the reactions generally becoming less efficient with higher dark matter masses.}
    \label{fig:decaykspace}
\end{figure}

In \fig \ref{fig:decaykspace}, we show the effect of these two scenarios on $\kDetach$ and $\kDiss$ for reference halos of mass $M_{200}=10^{10} M_\odot$ , Einasto shape of $\alpha=0.5$, and concentration parameters of $c=2.5$ and $c=20$. We also show the critical curves discussed in Sec. \ref{sec:CritCurve} for three different self-shielding parameterisations; direct collapse is possible where the dark matter curves are higher than the critical curves. Here, the dark matter lifetime is fixed at $\tau = 3\times10^{23}$~s and the mass ranges from $13.6$~eV to $100$~eV. The upper end of the dark matter curves corresponds to $m_{dm} = 13.6$ eV. The chemical reaction rates decrease for larger masses because a larger fraction of the produced photon spectrum is at energies above the LW range. Both the box and parabola spectra result in ratios of $\kDiss$ and $\kDetach$ comparable to blackbody spectra with temperatures in the range of $10^4$-$10^5$ K, showing that both examples can lead to similar outcomes as UV radiation from nearby star forming regions. The parabola spectrum peaks at an energy of $E=m_{dm}/2$. For masses where that peak does not sit within the LW range, this leads to a suppression in photodetachment and photodissociation. Since the two scenarios yield similar results, and since realistic decays are more similar to the parabolic shape, we will conservatively focus on results with this spectrum going forward. 

\subsubsection{Annihilation}
Relic dark matter may annihilate if it is self-conjugate, or if sufficient quantities of its conjugate particle are present. Since the annihilation rate then scales with the square of the density, annihilation will yield a stronger environmental dependence relative to the decay scenario. For an $s$-wave annihilating dark matter component with mass $m_{dm}$ and thermally averaged annihilation cross section of $\sigv$, the photon injection rate is given by
\begin{equation}
    \frac{dn_\gamma}{dEdt}(E,r) = \frac{f_\gamma \fdm^2\rho_{dm}^2(r) \sigv }{m_{dm}^2 2^p} \frac{dN}{dE}(E),
\end{equation}
where $p=1$ for self-conjugate dark matter (e.g. a Majorana fermion) and $p=2$ for dark matter which consists of a particle and antiparticle (e.g. Dirac).

The photon flux \eqref{eq:centreFlux} can again be computed analytically:
\begin{equation}
    J(E) = \frac{f_\gamma \fdm^2 \sigv}{m_{dm}^2 2^p} E \frac{dN}{dE}(E) \rho_0^2 r_0 g_{ann}(\alpha),
\end{equation}
where 
\begin{equation}
    g_{ann}(\alpha)\equiv\bigg(\frac{e^4 \alpha}{4}\bigg)^{1/\alpha} \Gamma\bigg(1+1/\alpha\bigg) = \bigg(\frac{e^2 }{2}\bigg)^{1/\alpha} g_{dec}(\alpha).
\end{equation}
We again focus on annihilations to more than two final-state particles in order to obtain a continuum of photons, and use the parabola spectrum from \eq \eqref{eq:parabolaSpectrum}. As is the case for decaying dark matter, the spectral shape of outgoing photons affects the $\kDiss$ and $\kDetach$ ratio. For a given dark matter mass and spectral shape, the ratio of the rates of the two chemical reactions will be the same for annihilating and decaying dark matter. Changes in the annihilation cross section affect the overall intensity of the flux scaling both $\kDiss$ and $\kDetach$ by the same factor so that the ratios will remain the same as shown in \fig \ref{fig:decaykspace}.

We will examine both thermal and non-thermal dark matter scenarios. For both scenarios we use a reference case of a self-conjugate particle where all of the annihilation products are photons corresponding to $p=f_{\gamma} = 1$. Thermally produced dark matter has annihilation cross sections of 
\begin{equation}
    \sigv \ge \sigv_{th} \approx 3\times 10^{26} \textrm{ cm}^3/\textrm{s},
\end{equation} 
where the equality would saturate the observed relic density. 
Sub-MeV thermally-produced dark matter cannot comprise all of the dark matter in the Universe \cite{Boehm:2013jpa}. However larger annihilation cross sections yield smaller relic abundances such that
\begin{equation} \label{eq:fdmThermal}
    \fdm = \frac{\sigv_{th}}{\sigv}.
\end{equation}
Therefore the thermally produced dark matter relevant for DCBHs can be a small subcomponent of the total dark matter in the Universe.

Non-thermal annihilating dark matter must have a small enough interaction rate to avoid thermalisation with the hot plasma in the early Universe. If it is produced ``cold'', it would  sidestep any bounds on extra radiation during nucleosynthesis or recombination. Freeze-in is not necessarily a concern in this scenario, since the reverse annihilation process would require annihilation of $n > 2$ particles into dark matter. For the case of non-thermal dark matter we assume $\fdm=1$ and study annihilation cross sections of $\sigv < \sigv_{th}$.
Typically for two-to-two particle annihilation, a $\sigv$ smaller but similar to $\sigv_{th}$ results in overabundant dark matter. However, due to the asymmetric nature of the annihilation processes in this work, even for $\sigv \sim \sigv_{th}$ thermal production in the early Universe could be strongly suppressed.

\section{Constraints and Results} \label{sec:results}

Using the above derivations of the dependence of the produced UV photon flux on the dark matter particle and halo models, we are able to identify the parameter space of dark matter models where $\mathrm{H}_2$ is suppressed and DCBHs can form. The viable parameter space is limited by indirect searches for decaying and annihilating dark matter. In this section we compare the region of interest to the existing constraints for both classes of dark matter energy injection and we will demonstrate that annihiating dark matter is able to produce DCBHs in typical halos but that under standard assumptions of halo mass and concentration, decaying models which would do so are ruled out by astrophysical observations.

\subsection{Decay}
Strong constraints exist on  decaying dark matter in the 10-100 eV range. The strongest constraint on dark matter that dominantly decays to LW and lower energy photons comes from anisotropies in the Cosmic Optical Background~(COB)~\cite{Nakayama:2022jza}. The COB constraints were produced assuming dark matter decays to two photons with energy $E_\gamma = m_{dm}/2$. 

We recast the COB constraints for decays to extended photon spectra. That is done by assuming that for extended decay spectra, the sum over all energies of the ratio of photon production rate at a given energy to the limit on the rate of photon production at that energy in the monoenergetic decay case must be less or equal to 1. Mathematically we require,
\begin{equation}
    \int dE \frac{\Gamma \frac{dN}{dE}}{2\Gamma_{2\gamma}(m_{dm}=2 E)} < 1,
\end{equation}
where $\Gamma = 1/\tau$ is the decay rate and $\Gamma_{2\gamma}(m_{dm})$ is the limit on dark matter of mass $m_{dm}$ decaying to two photons with energy $m_{dm}/2$. A 95\% confidence-level limit on the total lifetime $\tau$ from the COB can therefore be found to be
\begin{equation} \label{eq:tauCOB}
    \tau > \int dE \frac{\tau_{2\gamma}(m_{dm}=2E)}{2} \frac{dN}{dE} .
\end{equation}

CMB constraints on the ionization floor during the dark ages also apply. Using {\it Planck} \cite{Planck:2015fie} data, Ref.  \cite{Slatyer:2016qyl} obtained the 95\% confidence-level constraint:
\begin{equation} \label{eq:tauReion}
    \tau \ge \fdm f_\textrm{eff} 2.6\times 10^{25}~\mathrm{s} .
\end{equation}
The fraction of photon energy contributing to ionizing the gas is characterized by $f_\textrm{eff}$. For very light decaying dark matter we can assume all photons with $E>13.6$ eV are absorbed and all lower energy photons are not. This implies that
\begin{equation} \label{eq:feff}
    f_\textrm{eff} = \frac{1}{m_{dm}}\int_{13.6\textrm{ eV}}^{m_{dm}} E\frac{dN}{dE}dE  .
\end{equation}

\begin{figure}[htb]
    \centering
    \includegraphics{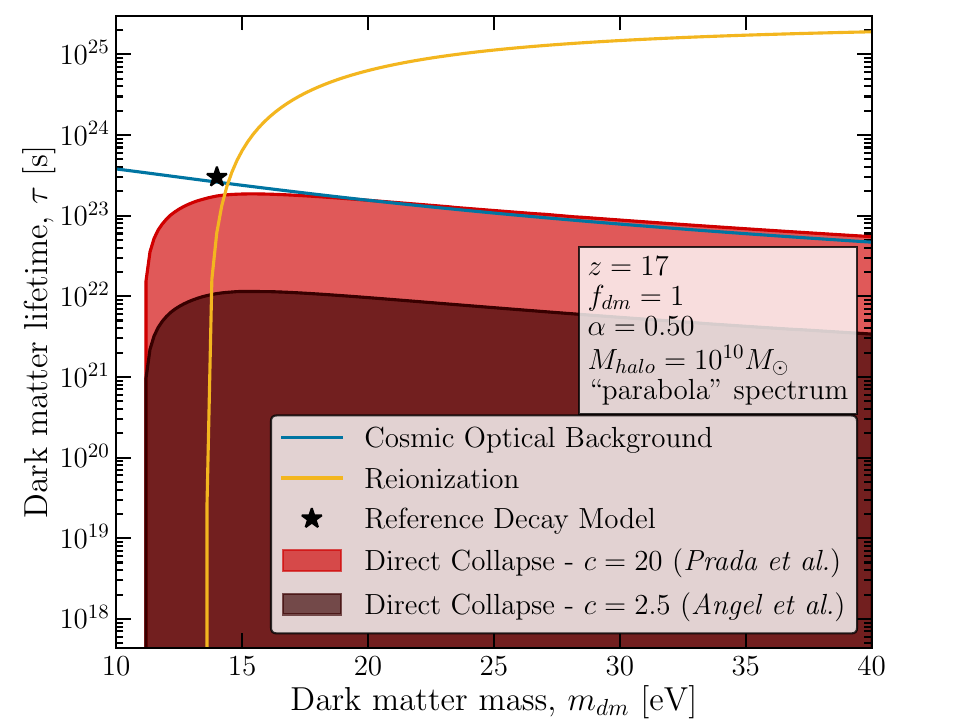}
    \caption{Parameter space of dark matter that decays to a parabola spectrum of photons which is able to produce DCBHs in a $10^{10} M_\odot$ mass halo at redshift $z=17$. The \pradaColour shaded region corresponds to models that suppress $\mathrm{H}_2$ in halos with concentration that follows the {\it Prada et al.} $c-M$ relationship \cite{Prada2012} whereas the \angelColour region uses the {\it Angel et al.} $c-M$ relationship \cite{Angel:2015ilq}. The halo is assumed to have a shape parameter of $\alpha=0.5$, inline with the best fit from Ref. \cite{Gao:2007gh}. The \cobColour curve shows the lower limit on dark matter lifetime from Cosmic Optical Background anisotrpies recast from Ref. \cite{Nakayama:2022jza} as expressed in \eq \eqref{eq:tauCOB}. The \reionColour curve shows the lower limit on dark matter lifetime from {\it Planck 2015} reionization observations \cite{Planck:2015fie} as described in \eq \eqref{eq:tauReion}. Lifetime constraints represent 95\% confidence-level exclusion curves. The black star shows the dark matter model which is used as a reference in \Fig \ref{fig:decayCM}.}
    \label{fig:decayMdmTau}
\end{figure}

\begin{figure}[htb]
    \centering
    \includegraphics[width=0.48\textwidth]{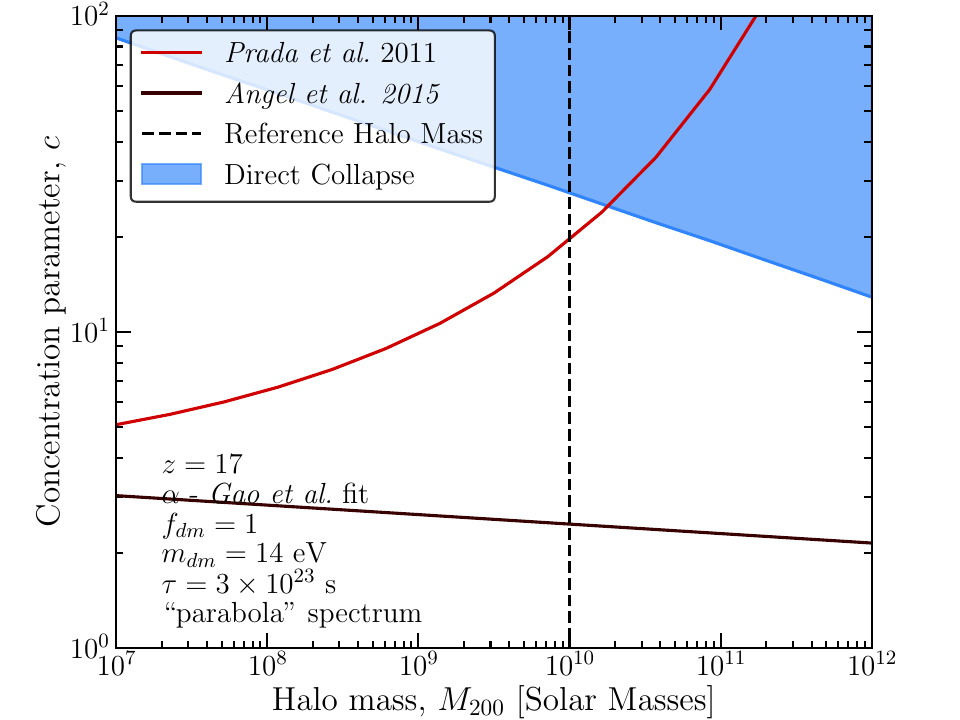}
    \includegraphics[width=0.48\textwidth]{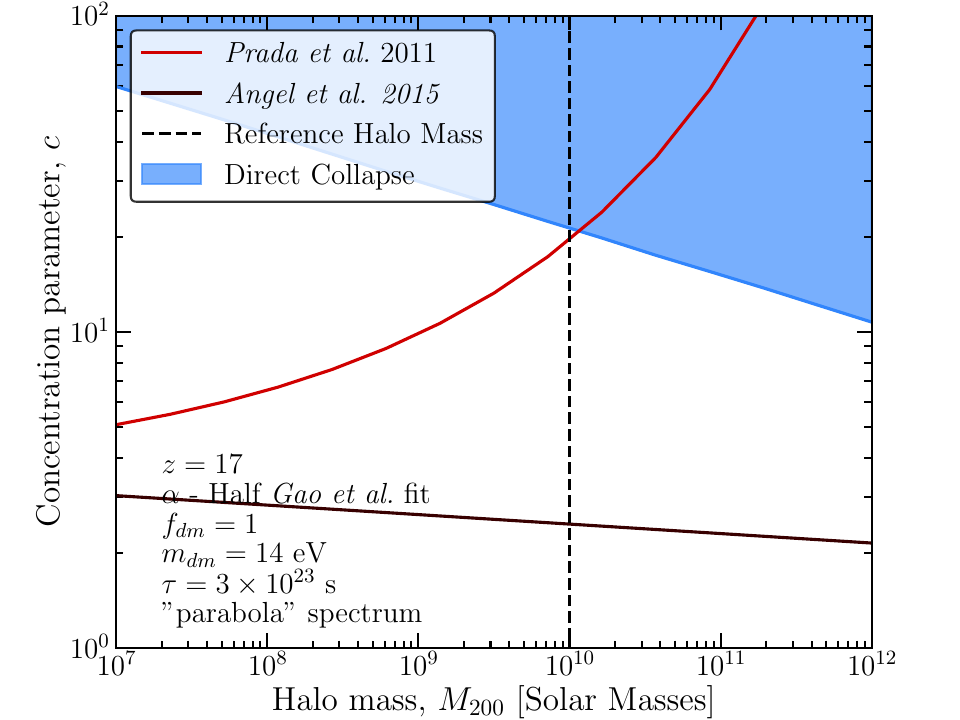}
    \caption{Range of halo concentration, $c$, and mass, $M_{200}$, values that can produce DCBH if dark matter is completely comprised of a 14 eV mass particle which decays to a parabola spectrum of photons with a lifetime of $3\times10^{23}$ s. This model corresponds to the black star in \fig \ref{fig:decayMdmTau}. The \haloRegionColour shaded regions show halos where direct collapse can occur while the \pradaColour curve shows the $c-M$ relation from {\it Prada et al.} \cite{Prada2012} and \angelColour curve shows the same relation from {\it Angel et al.} \cite{Angel:2015ilq}. The black dashed lines show reference Halo mass of $10^{10} M_\odot$. Left: Einasto $\alpha$ parameter represents average halo cuspiness according to Ref. \cite{Gao:2007gh}. Right: Einasto $\alpha$ parameter is half of the average value from Ref. \cite{Gao:2007gh}.}
    \label{fig:decayCM}
\end{figure}

Figure \ref{fig:decayMdmTau} shows dark matter parameter space where decays to a parabola spectrum of photons in a $10^{10}\; M_\odot$ mass halo at redshift $z=17$ with an Einasto shape parameter of $\alpha=0.5$ would produce a UV backgorund that is expected to suppress $\mathrm{H}_2$ formation. The red shaded regions indicates the decaying dark matter parameter space where the produced UV flux is above the {\it Luo et al.}~\cite{Luo_2020} critical curve obtained using a self-shielding characteristic length of $L_\textrm{Jeans}/4$. The \pradaColour region uses {\it Prada et al.}'s~\cite{Prada2012} mass-concentration relation, whereas the \angelColour region shows the more conservative parameter space using the {\it Angel et al.}~\cite{Angel:2015ilq} results. The \cobColour and \reionColour curves respectively show lower limits on dark matter lifetime coming from COB anisotropies and reionization as described by \eq \eqref{eq:tauCOB} and \eq \eqref{eq:tauReion}. 

All decaying dark matter models which are able to produce DCBHs in these reference halos are ruled out by observations. Fig. \ref{fig:decayCM} shows the halo parameter space where $\mathrm{H}_2$ abundance would be suppressed for a viable dark matter model with mass $m_{dm} = 14$ eV and lifetime $\tau = 3\times 10^{23}$ s. This reference dark matter model is noted by the black star in \fig \ref{fig:decayMdmTau}. The left panel of Fig. \ref{fig:decayCM} shows halos with Einasto $\alpha$ parameters following the average from the {\it Gao et al.} fit \cite{Gao:2007gh}, while the plot on the right shows cuspier halos where $\alpha$ is half of the predicted average value. Even for highly concentrated halos that follow the {\it Prada et al.} $c$-$M$ relationship, halos with masses of $10^{10} M_\odot$ would need to have $\alpha$ values about half the average halo to produce a DCBH. Moving to higher halo masses, $M_{200}=3\times10^{10}M_\odot$, we see that direct collapse can occur for an average $\alpha$ parameter, but only for the optimistic \textit{Prada}-like mass-concentration relation. This highlights the importance of quantifying the expected abundance and diversity of large, concentrated, and cuspy halos to determine whether decaying dark matter could produce DCBHs in enough halos to account for the high-$z$ quasar population.

\subsection{Annihilation}

Due to the model building challenges to be discussed in Section \ref{sec:modelBuilding}, there are not many constraints on very light dark matter models which annihilate to a spectrum of photons. The annihilation rate is again constrained by the effects dark age energy injection on the CMB. This is parameterised via:
\begin{equation}
    \pann \equiv f_\textrm{eff} \fdm^2 \frac{\sigv}{m_{dm}}
\end{equation}
where $f_\textrm{eff}$ is the fraction of annihilation energy contributing to IGM ionisation or heating as described in \eq \eqref{eq:feff}. $\pann$ must be below the observational upper limit, $\pmax$. {\it Planck 2018} found the 95\% confidence-level upper limit to be \cite{Planck:2018vyg}
\begin{equation}
    \pmax = 3.2 \times 10^{-37} \textrm{cm}^3/\textrm{s}/\textrm{eV} \ge \pann .
\end{equation}

For non-thermal dark matter this sets a thermally-averaged cross section upper limit
\begin{equation} \label{eq:upperNTsigv}
    \sigv \le 3.2\times 10^{-36}\textrm{cm}^3/\textrm{s }  \bigg(\frac{1}{f_\textrm{eff}}\bigg) \bigg(\frac{m_{dm}}{10 \textrm{ eV}}\bigg) \bigg(\frac{1}{\fdm}\bigg)^2 .
\end{equation}
In the case of thermally produced dark matter, reionization sets a lower bound on the annihilation cross section because as $\sigv$ gets smaller, the abundance of the annihilating dark matter grows. Using \eq \eqref{eq:fdmThermal} to remove the explicit $\fdm$ dependence in \eq \eqref{eq:upperNTsigv} yields the constraint on thermal dark matter's annihilation cross section
\begin{equation} \label{eq:lowerTHsigv}
    \sigv \ge 2.8\times 10^{-16} \textrm{cm}^3/\textrm{s } f_\textrm{eff} \bigg(\frac{10 \textrm{ eV}}{m_{dm}}\bigg)  \bigg(\frac{\sigv_{th}}{3\times 10^{-26} \textrm{cm}^3/\textrm{s }}\bigg)^2 .
\end{equation}
This can equivalently be described as an upper limit on the fraction of dark matter comprised of this light annihilating subcomponent
\begin{equation} \label{eq:fdmUpperLim}
    \fdm \le 1.1\times10^{-10} \bigg(\frac{1}{f_\textrm{eff}}\bigg) \bigg(\frac{m_{dm}}{10 \textrm{ eV}}\bigg)  \bigg(\frac{3\times 10^{-26} \textrm{cm}^3/\textrm{s }}{\sigv_{th}}\bigg).
\end{equation}
Therefore any light thermal dark matter candidate of interest to DCBH production will be a very small subcomponent of the overall dark matter abundance of the Universe.

\begin{figure}[htb]
    \centering
    \includegraphics[width=0.48\textwidth]{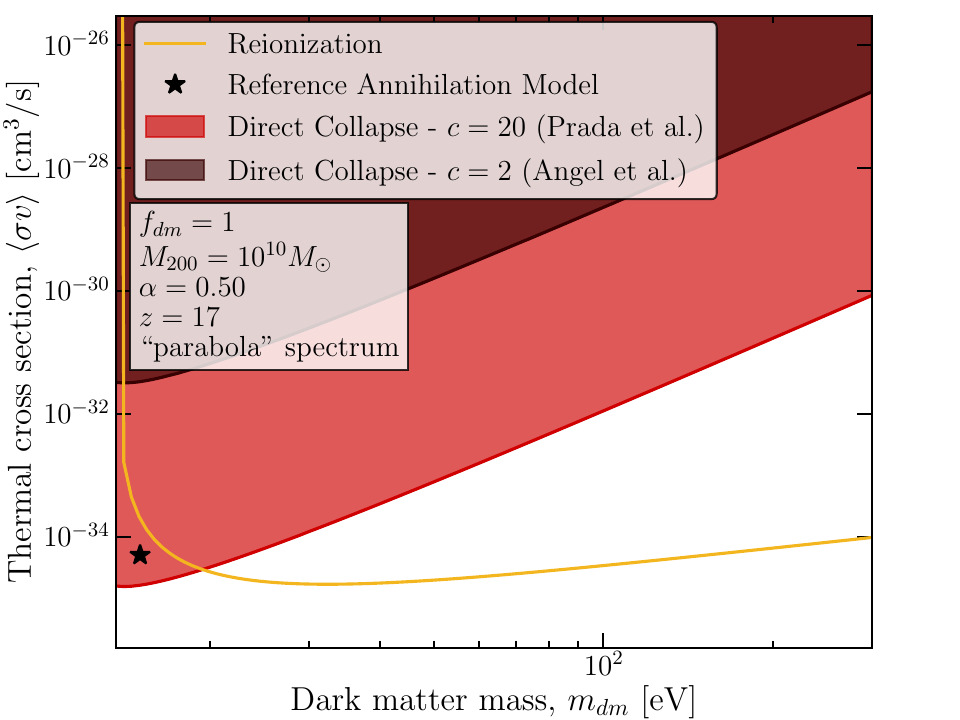}
    \includegraphics[width=0.48\textwidth]{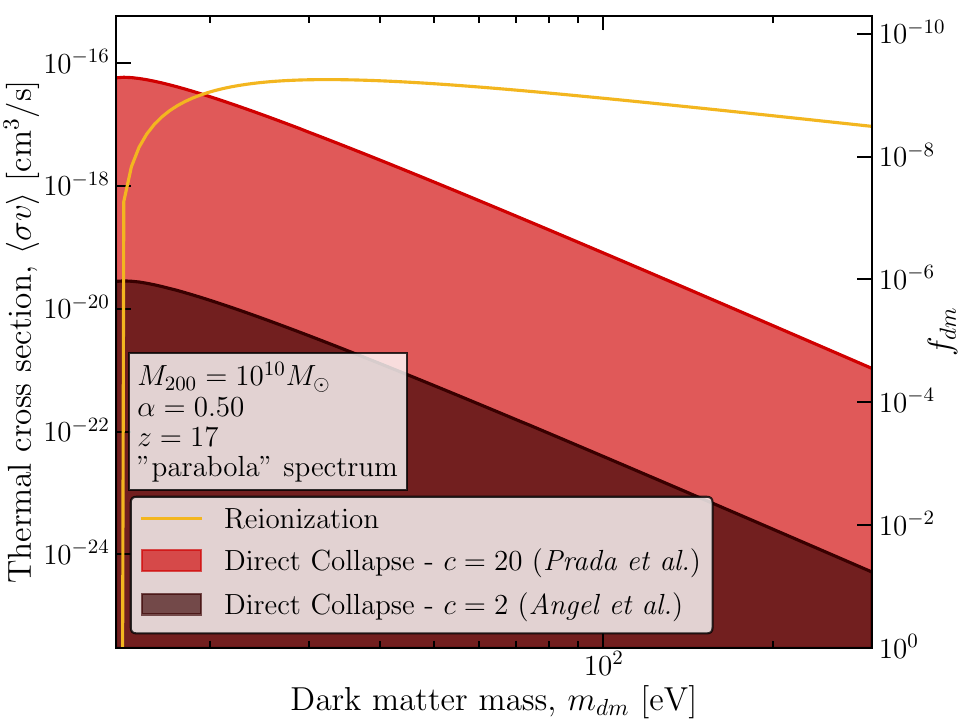}
    \caption{Parameter space of self-conjugate dark matter that annihilates to a parabola spectrum of photons which is able to produce DCBHs in a $10^{10} M_\odot$ mass halo at redshift $z=17$. The \pradaColour shaded region corresponds to models that suppress $\mathrm{H}_2$ in halos with concentrations which follow the {\it Prada et al.} $c-M$ relationship \cite{Prada2012} whereas the \angelColour region uses the {\it Angel et al.} $c-M$ relationship \cite{Angel:2015ilq}. The halo is assumed to have a shape parameter of $\alpha=0.5$, inline with the best fit from Ref. \cite{Gao:2007gh}. The \reionColour curves depict 95\% confidence-level limits on the theramlly-averaged annihilation cross section, $\sigv$, from {\it Planck 2018} constraints on the maximum power of ionizing radiation during the dark ages \cite{Planck:2018vyg}. Left: Parameter space for non-thermal dark matter that comprises all of the dark matter in the Universe. The \reionColour curve is an upper bound on $\sigv$. The black star shows the dark matter model which is used as a reference in \Fig \ref{fig:annCM}. Right: Parameter space for thermally produced dark matter that comprises a subcomponent of dark matter with a fraction of $\fdm = \frac{\sigv_{th}}{\sigv}$. Here to avoid constraints, dark matter models must be above the \reionColour curve with a larger $\sigv$ and a smaller $\fdm$.}
    \label{fig:annDM}
\end{figure}

\begin{figure}[htb]
    \centering
    \includegraphics[width=\textwidth]{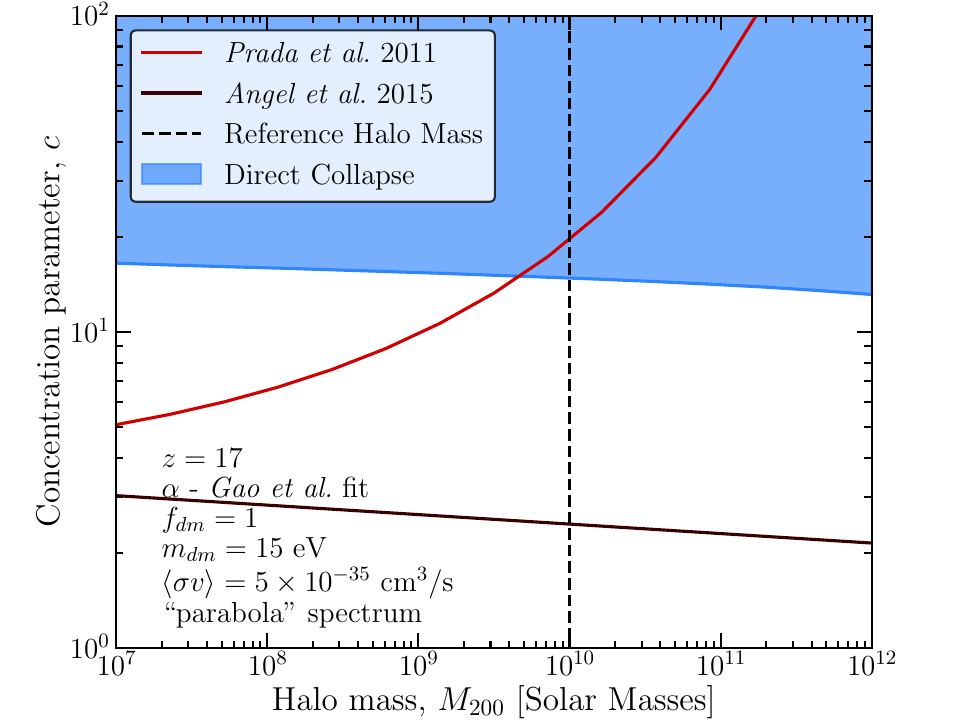}
    \caption{Range of halo concentration, $c$, and mass, $M_{200}$, values that can produce DCBHs if dark matter is completely comprised of a 15 eV mass self-conjugate particle which annihilates to a parabola spectrum of photons with a thermally-averaged cross section of $5\times10^{-35}$ cm$^3/$s. This model corresponds to the black star in \fig \ref{fig:annDM}. The \haloRegionColour shaded region shows halos where direct collapse can occur while the \pradaColour curve shows the $c-M$ relation from {\it Prada et al.} \cite{Prada2012} and \angelColour curve shows the same relation from {\it Angel et al.} \cite{Angel:2015ilq}. The black dashed lines show reference halo mass of $10^{10} M_\odot$ and for all halo masses, the Einasto shape parameter, $\alpha$, is assumed to be the average value according the {\it Gao et al.} fit \cite{Gao:2007gh}.}
    \label{fig:annCM}
\end{figure}

\fig \ref{fig:annDM} shows the annihilating dark matter parameter space able to suppress $\mathrm{H}_2$ formation in a reference halo with mass $M_{200}=10^{10}M_\odot$. The left panel shows the case of non-thermal self-conjugate dark matter with $f_{dm}=1$ while the right panel shows thermally-produced dark matter where the fraction of dark matter comprised of this light annihilating subcomponent is given by \eq \eqref{eq:fdmThermal}. The \reionColour curve shows the limits on $\sigv$ from reionization as described in \eqs \eqref{eq:upperNTsigv}~and~\eqref{eq:lowerTHsigv}. The \pradaColour shaded region corresponds to dark matter models that can produce DCBHs with concentrations of $c=20$, following {\it Prada et al.} \cite{Prada2012} whereas the \angelColour region shows the dark matter models that still produce DCBHs in less concentrated halos following the {\it Angel et al.} $c-M$ relationship \cite{Angel:2015ilq}.

For both non-thermal and thermal dark matter, there is a region of parameter space that has not been excluded by reionization observations which are able to produce DCBHs in halos with large but realistic concentrations. Using the {\it Prada et al.} $c$-$M$ relationship, this parameter space corresponds to dark matter in the mass range of $13.6 \textrm{ eV} \le m_{dm} \le 20 \textrm{ eV}$. In the case of non-thermal dark matter the thermally averaged cross section would be on the order of $\sigv \sim 10^{-35}$ cm$^3/$s. Alternatively when the annihilating dark matter is thermally produced the thermally-averaged cross section would be on the order of $\sigv \sim 10^{-17}$ cm$^3$/s and comprise a fraction of about $10^{-9}$ of the total dark matter abundance.

The blue shaded region in Fig.~\ref{fig:annCM} shows the halo mass and concentration parameter space which would suppress molecular hydrogen formation allowing for DCBHs to potentially form. This is shown for a non-thermal annihilating dark matter model, depicted as a black star in \fig \ref{fig:annDM}, which has not been excluded by {\it Planck}. The \pradaColour and \angelColour lines show the $c$-$M$ relationships from {\it Prada et al.} \cite{Prada2012} and {\it Angel et al.} \cite{Angel:2015ilq} respectively. If real halos have concentrations that are more accurately described by the {\it Prada et al.} results, then an annihilating dark matter model with mass $m_{dm}=15$ eV and $\sigv=5\times10^{-35}$ cm$^3$s$^{-1}$ which comprises all of the dark matter in the Universe would be able to produce large black hole seeds in halos with masses even lower than the reference halos mass of $10^{10}M_\odot$. This demonstrates that dark matter annihilating to an extended spectrum of photons is a viable method for producing DCBHs at redshifts of $z=17$.

%----------------------------

\section{Discussion} \label{sec:discussion}
\subsection{Validity of comparison to critical curve} \label{sec:critCurveValid}
The critical curves used to define the necessary UV flux for direct collapse, and the calculation of the flux produced by dark matter use different models and a different set of assumptions. Combining these two methods allows for a proof-of-concept demonstration that dark matter energy injection can create conditions needed for direct collapse to occur. However, a fully self-consistent study of dark matter energy injection along with the collapse and chemical dynamics could produce different results, potentially even showing direct collapse to be more easily explained by light decaying or annihilating dark matter.

There are two primary differences between the assumptions behind the critical curve produced by {\it Luo et al.} \cite{Luo_2020} and the energy injection calculation performed in this work. First, the critical curve calculation assumes that the UV source is from a nearby star-forming region outside of the gas cloud, whereas injection from dark matter leads to in-situ UV production. This difference could have important effects on the uniformity of the UV flux and the validity of the $\mathrm{H}_2$ self-shielding approximation. Second, the critical curve was determined using a 3D hydrodynamic code, whereas we have focused on injection in the central part of the halo. Spatial nonuniformity of the flux produced by dark matter energy injection could affect the overall collapse rate.

In the traditional scenario of UV photons originating in a nearby star forming region, the collapsing gas cloud sits within a homogeneous but anisotropic photon bath. The only source of spatial inhomogeneity is the $\mathrm{H}_2$ self-shielding such that the inner parts of the gas cloud experience a reduced flux once the gas becomes optically thick. Additionally, the anisotropy of the photon bath affects the direct collapse process as the side of the gas cloud farther from the nearby galaxy receives a smaller UV flux due to the increased $H_2$ self-shielding \cite{Regan2016} however this effect is not accounted for by \textit{Luo et al.} when obtaining the critical curves used in this work \cite{Luo_2020}. When the UV photons are produced from the cloud's dark matter halo the flux will be nearly isotropic, more closely aligning with the assumption used in the critical curve derivation than the traditional direct collapse scenario. However, in the optically thin regime the flux produced by dark matter will be highly inhomogeneous as it will be largest close to the halo centre and reduce at larger radii. 

In this work we have determined the photodetachment and photodissociation rates at the centre of the halo with the assumption that that is the most important location to study because if the gas cloud and halo are nearly cocentric this is where the black hole is expected to form. However, it is likely that even in scenarios where $\mathrm{H}_2$ is suppressed in the centre of the gas cloud, there would still be outer regions of the gas cloud where $\mathrm{H}_2$ can easily form, leading to efficient cooling and fragmentation around the more slowly-contracting core. While focusing on the central gas is a strong assumption, if efficient cooling in the outer region of the gas cloud is important then direct collapse may require a larger flux than we have assumed in this work. Properly understanding the impact of fragmentation in the outer halo requires embedding the dark matter energy injection consistently within a 3D hydrodynmic simulation.

In-situ photon injection also affects the geometry of $\mathrm{H}_2$ self-shielding. Throughout Section~\ref{sec:results} we have presented results using a self-shielding approximation where the characteristic length in \eq \eqref{eq:ColumnDensity} is one quarter of the Jeans length. {\it Luo et al.} found that  this approximation most accurately reproduced the column density in the traditional direct collapse scenario \cite{Luo_2020}. However, when the UV photons originate from inside the gas cloud there are two expected differences: 1) the average column density between the produced photon and a given point is reduced and 2) the inner regions experience less self-shielding whereas the outer regions experience more because most of the photons are produced in the dense central portion of the gas cloud.

\begin{figure}[htb]
    \centering
    \includegraphics[width=0.48\textwidth]{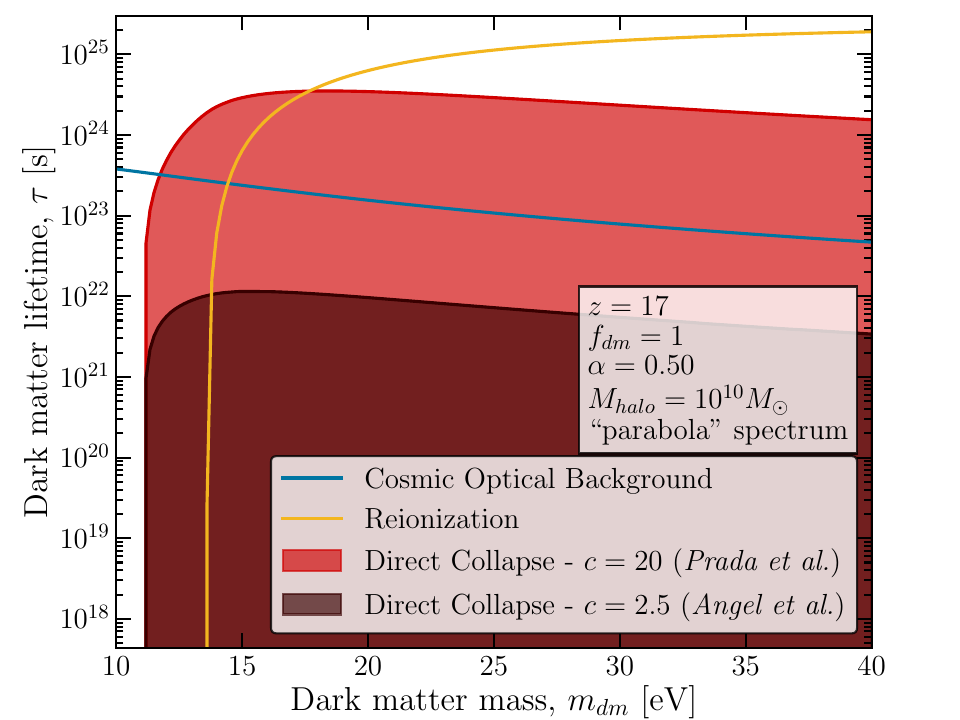}
    \includegraphics[width=0.48\textwidth]{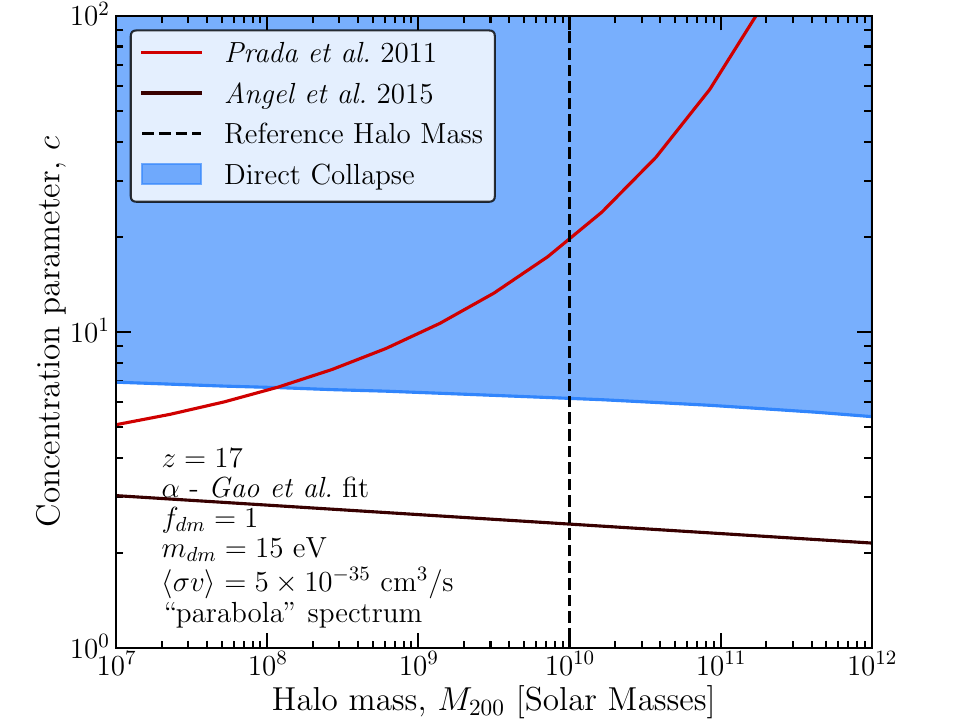}
    \includegraphics[width=0.48\textwidth]{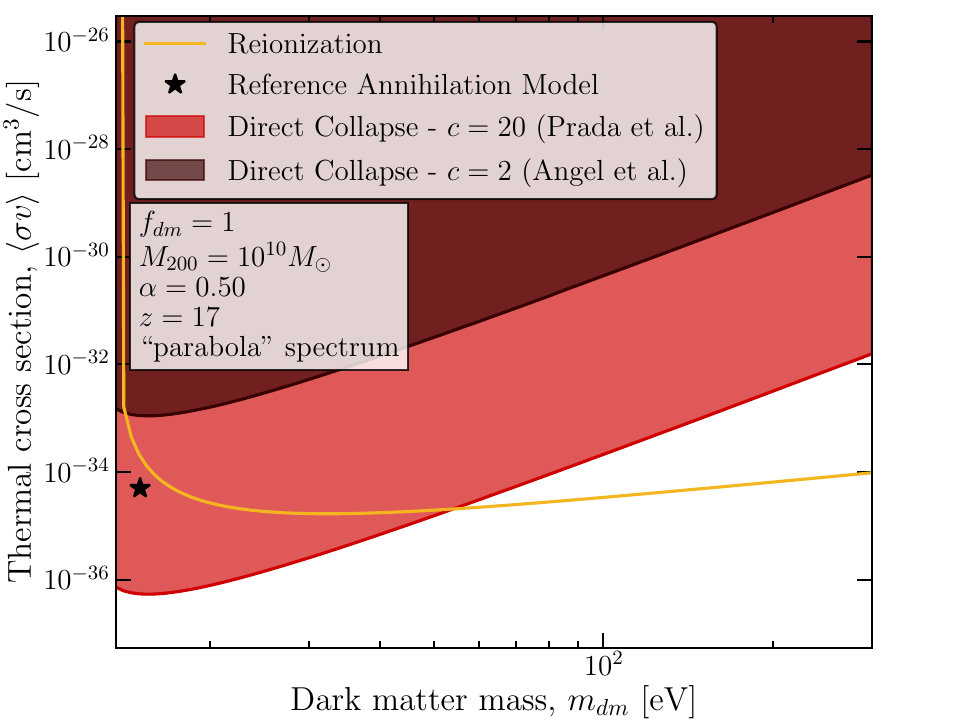}
    \includegraphics[width=0.48\textwidth]{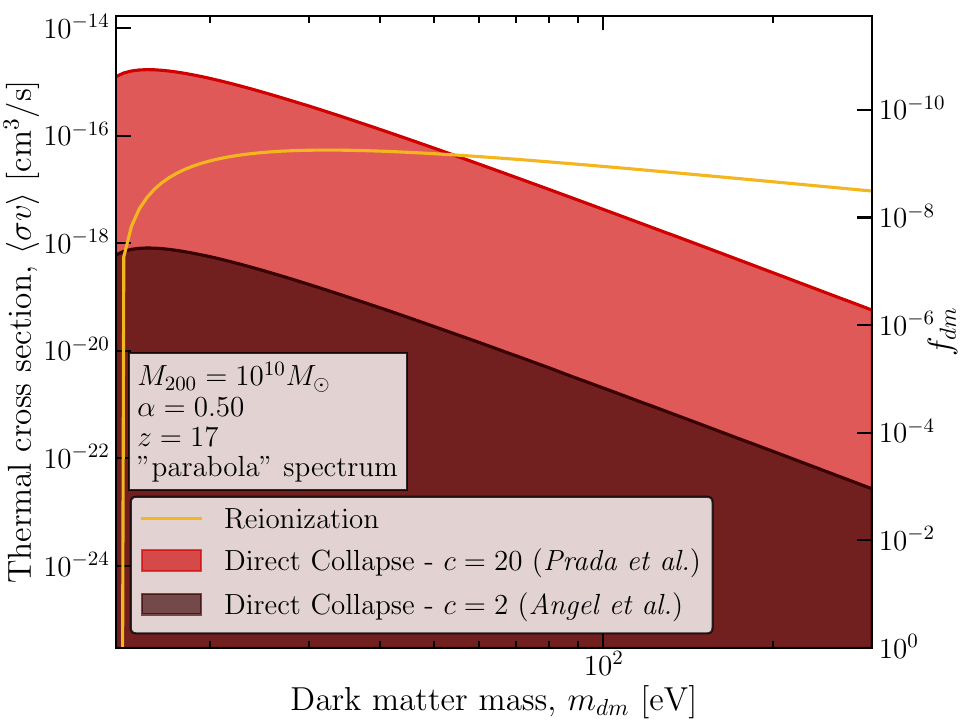}
    \caption{Summary of results shown in Section \ref{sec:results} except using the direct collapse critical curve that assumes a Sobolev-like characteristic length in self-shielding approximation of \eq \eqref{eq:ColumnDensity}. Top left: decaying dark matter parameter space previously described in \fig \ref{fig:decayMdmTau}. Bottom: non-thermal and thermal self-conjugate annihilating dark matter parameter space as in \fig \ref{fig:annDM}. The black star indicates the dark matter model which is shown in the top right panel. Top right: parameter space of halos where non-thermal annihilating dark matter can produce DCBHs as in \fig \ref{fig:annCM}.}
    \label{fig:sobolev}
\end{figure}

To illustrate the possible effect of reduced self-shielding, we summarise in \Fig \ref{fig:sobolev}  how the key results in Section \ref{sec:results} would change if the Sobolev-like self-shielding critical curve from {\it Luo et al.} is used instead of $L_{\rm Jeans}/4$. The top left panel in \fig \ref{fig:sobolev} shows the parameter space of dark matter which decays to a parabola spectrum of photons that could produce DCBHs. In contrast to \fig \ref{fig:decayMdmTau}, which shows the same parameter space with the quarter Jeans length self-shielding, there are unconstrained decaying dark matter models which are able to suppress $\mathrm{H}_2$ formation in halos which follow the {\it Prada et al.}~\cite{Prada2012} $c$-$M$ relationship. The dark matter models where that is true are shaded \pradaColour. COB and CMB (reionisation) bounds are the same as in \fig \ref{fig:decayMdmTau}.
The lower panels in \fig \ref{fig:sobolev} demonstrate the impact of reduced self-shielding on the relevant parameter space of non-thermal and thermal annihilating dark matter models which were previously shown in \fig \ref{fig:annDM}. When self-shielding is reduced, for {\it Prada et al.} concentrations, the range of viable models is extended up to $m_{dm} \sim55$ eV. The top-right panel of \fig \ref{fig:sobolev} shows the halo parameter space for the same dark matter model as \fig \ref{fig:annCM}. The shaded \haloRegionColour region shows the halos where $\mathrm{H}_2$ would be suppressed for a non-thermally produced dark matter model with $m_{dm}=15$ eV and $\sigv=5\times10^{-35}$ cm$^3$s$^{-1}$ if the Sobolev-like self-shielding approximation is used. For halos with the reference mass of $M_{200}=10^{10}M_\odot$, halos with concentrations as small as $c=4$ would produce a sufficient UV flux to form DCBHs. This concentration is still larger than the average predicted by {\it Angel et al.} \cite{Angel:2015ilq} as depicted by the \angelColour line. However, the required halo concentration is much lower than the prediction of {\it Prada et al.} \cite{Prada2012}, shown in \pradaColour. Additionally if the {\it Prada et al.} $c$-$M$ relationship does accurately describe the average halos in our Universe, molecular cooling could be suppressed in much smaller halos with masses of $M_{200} \gtrsim 10^8 M_\odot$. The changes depicted in \fig \ref{fig:sobolev} highlight the importance of  $\mathrm{H}_2$ self-shielding.

A final difference between the dark matter energy injection in this work and the setup used to determine the direct collapse critical curves is the production of ionising radiation. The presence of ionising radiation with energies $E>13.6$ eV can act as a catalyst for $\mathrm{H}_2$ formation, therefore requiring a larger UV flux to compensate and suppress molecular cooling \cite{Inayoshi_2011}. The critical curves used in this work do not account for such an X-ray background. However, both the parabola and the box spectrum shapes used here would lead to a sizable amount of ionising radiation when $m_{dm} \gg 13.6$ eV. Estimating the impact of ionisation would require a recalculating of the critical curves that we have used here. At most, however, this would lead to an upper limit on the energies produced in the decay or annihilation process for which dark matter-induced H$_2$ suppression is viable. This does not necessarily mean a lower dark matter mass, as partial decays or annihilations to heavy invisible states, for example, can yield such a kinematic limit. A precise understanding of how  an ionising background affects direct collapse in the dark matter energy injection scenario is therefore important for determining which dark matter models are viable, but would be unlikely to change the qualitative conclusions in this work.

As has been previously alluded to, we make heavy use of a simplified one-zone model to capture essential aspects of the direct collapse problem at manageable computational cost. While valuable insights can be gained from this approach, the evolution of hydrodynamic systems
%, even only constituent of primordial gas in isolated halos, 
play host to additional physical processes that are not included in this model, yet could impact the viability of direct collapse. In future work, we hope to more fully integrate these properties via a full hydro-simulation. Besides the markedly higher computational cost, this approach brings with it additional challenges, in particular the resolution of a very high range of densities over a halo sized object. Nonetheless, works by \cite{Shang_2010, Schauer2017, Luo_2020, Latif2022nat} have demonstrated how direct collapse can be modelled in simulations, both in the context of a cosmological box and isolated halo zoom in simulation. The introduction of decay and annihilation products would essentially act as an additional fluid or radiation field, with their precise treatment somewhat dependent on the dark matter model. In the case of low energy photons studied in this work, which can be treated as being absorbed by the gas largely in situ, numerical treatment akin to that of standard astrophysical radiation fields like the cosmic UV background may be appropriate. In contrast when considering the annihilation products of more massive dark matter particles, one must take into account the production of particle cascades and energy transfer beyond a single gas cell \cite{Schon2018}. Some of this can be done via existing ray tracing and radiative transfer methods such as the existing treatment (for example \cite{Hopkins2019}) of cosmic rays and resolved stellar particles. However these methods may become too numerically expensive in the case of dark matter decay and annihilation and some kind of semi-analytic compromise may be called for. For now we leave these quandaries for future consideration. 

\subsection{Dark matter models} \label{sec:modelBuilding}
We have demonstrated that dark matter decays and annihilations are able to produce a UV background capable of suppressing $\mathrm{H}_2$ formation in early gas clouds. However, the decaying and annihilating dark matter models that we have focused on have key differences to models most commonly discussed in the literature. 

Beyond the requirement of three or more particles in the annihilation or decay final state, the low mass of the dark matter models also raises model-building challenges. If the annihilating or decaying dark matter particle comprises all of the dark matter in the Universe, it cannot have been produced thermally. Observation of the Lyman-$\alpha$ forest from distant quasars has set constraints on thermally produced dark matter, requiring $m_{dm} > 4.09$ keV \cite{Baur:2015jsy}. The parameter space of thermally-produced annihilating dark matter we present has $\fdm \ll 1$, so these constraints do not apply. However, when building a model of the non-thermal annihilating or decaying dark matter, thermal production must be suppressed in the early universe and another production mechanism must exist. Additionally, regardless of the production mechanism, the Tremaine-Gunn bound limits the mass of fermionic dark matter when $\fdm \sim 1$~\cite{Tremaine:1979}. Observations of dwarf spheroidal galaxies have refined this limit to $m_{dm,\mathrm{Fermion}} \ge 190$ eV \cite{Savchenko:2019qnn}. This implies that in the scenarios where $\fdm \sim 1$, the dark matter particle should be bosonic, though more exotic scenarios such as a dark sector made up of a large number of similarly behaving species \cite{Davoudiasl:2020uig} may also be viable.

While the thermally annihilating dark matter scenario avoids the astrophysical mass constraints because $\fdm \ll 1$, the associated large annihilation cross sections presents model-building challenges. The dark matter models which suppress $\mathrm{H}_2$ and avoid constraints from the ionization floor have thermally-averaged annihilation cross sections on the order of $\sigv \sim 10^{-16}$ cm$^3$s$^{-1}$. This large cross section implies large effective couplings to photons. When constructing a fundamental model which includes such large effective couplings, constraints will likely arise from effects on atomic and molecular physics as well as e.g. stellar cooling bounds---though we may well find ourselves in the optically-thick regime where energy is reabsorbed in the star and these bounds do not apply. These constraints will be model-dependent and therefore do not directly factor into the parameter space explored in Sec. \ref{sec:results}.

Beyond the general classes of models discussed above, other scenarios such as excited dark matter \cite{Finkbeiner:2007kk} with a small mass splitting, Sommerfeld-enhanced \cite{Hisano:2004ds} or $p$-wave annihilating dark matter where the thermally-averaged cross section is temperature dependent, or extended dark sectors are capable of producing DCBHs. By constructing fundamental dark matter models that match onto the general classes discussed in this work or exploring the impact of different classes of dark matter models, the connection between particle physics and early structure formation can be better understood.  

%----------------------------

\section{Conclusions} \label{sec:conclusion}

In this work we have presented a new scenario for the monolithic collapse of gas clouds into large black hole seeds in the early Universe, which would evolve to power the large observed quasars at high redshifts. As a proof-of-concept, we have demonstrated that molecular hydrogen formation in early gas clouds can be suppressed by UV photons produced by dark matter within the gas cloud's halo. This scenario is viable for annihilating dark matter with a mass in the range of $13.6\textrm{ eV}\le m_{dm} \lesssim 20\textrm{ eV}$ which annihilates into an extended spectrum of photons. Such a candidate can be thermally produced with a large annihilation cross section, yielding a small fraction of the total dark matter content with $\fdm\sim10^{-9}$. Alternatively, the annihilating dark matter could produce a sufficient UV photon flux with a much smaller annihilation cross section, $\sigv \sim 10^{-35}$ cm$^{3}$s$^{-1}$, if it makes up all of the dark matter and was produced ``cold'' through some non-thermal mechanism.

We also examined the viability of models where dark matter decays to an extended spectrum of photons producing a sufficient UV flux to cause DCBHs. Unconstrained decaying dark matter models produce a smaller flux than their annihilating counterparts making it more difficult to explain the production of supermassive black holes. The energy injection from dark matter depends strongly on the properties of the halo itself and therefore decaying dark matter is able to suppress $\mathrm{H}_2$ formation in halos which are larger, cuspier, or more concentrated than what is currently predicted of the average halo at redshifts of $z=17$.

While the results presented here act as a compelling proof-of-concept, there remains work to be done to better understand the direct collapse process, the cosmological context of the halos, and particle physics in order to conclusively determine precisely which dark matter models could plausibly explain the observed quasar population. In Section \ref{sec:critCurveValid} we discuss the potential issues that arise from comparing the flux from dark matter energy injection within the host halo to direct collapse critical curves which were obtained assuming the UV flux originates from outside the gas cloud. Incorporating the dark matter energy injection directly into a 3D hydrodynamic simulation of the collapse and chemical network will provide a more accurate description of the impact that different dark matter models have on the collapse process. This should also help elucidate the impact of the spatial inhomogeneity of the produced UV flux as well as how to best model the $\mathrm{H}_2$ self-shielding. If the self-shielding is indeed reduced relative to direct collapse with an external UV source as we have hypothesized, then the main results shown in Section \ref{sec:results} will be overly conservative.

Further work improving and applying cosmological simulations will also be able to provide important insight into how well dark matter energy injection can explain the observed quasar population. Cosmological simulations are required to produce more precise understandings of the abundance of halos of different shapes and sizes in the early Universe as well as a quantitative result for the required frequency of halos undergoing direct collapse to reproduce the actual quasar population. That information is needed to be able to compare the impact of a given dark matter model with astrophysical observations. In this work we have shown how to determine if a specific dark matter halo comprised of a specific dark matter particle can undergo direct collapse but further work is needed before we are able to map a dark matter model onto a prediction for an observed quasar population. Additionally, cosmological simulations can be used to explore the impact of differing expansion histories and dark matter self-interactions. This will help determine whether consistent dark matter models exist which annihilate or decay to UV photons and also impact halo formation such that they are larger, cuspier, or more concentrated resulting in a larger than expected photon flux.  

We have shown that in halos consistent with cosmological simulations, light annihilating dark matter models are able to produce a sufficient UV flux while avoiding potential issues that exist in the traditional direct collapse scenario, such as metal contamination. The promising results shown in this work motivates the need for further exploration of light annihilating dark matter models as well as incorporating dark matter energy injection into future simulations of collapsing gas clouds.
    
%----------------------------
\acknowledgments
  AF is supported by an Ontario Graduate Scholarship. Funding for SS work was provided by the Charles E. Kaufman Foundation of the Pittsburgh Foundation. This work was supported at Pennsylvania State University by NASA ATP Program No. 80NSSC22K0819. ACV is supported by the Arthur B.~McDonald Canadian Astroparticle Physics Research Institute, NSERC, and the province of Ontario via an Early Researcher Award. Equipment is funded by the Canada Foundation for Innovation and the Province of Ontario, and housed at the Queen's Centre for Advanced Computing. Research at Perimeter Institute is supported by the Government of Canada through the Department of Innovation, Science, and Economic Development, and by the Province of Ontario. 
%----------------------------

\appendix
\section{Halo Concentration-Mass Fits} \label{sec:cMeqs}

Throughout this work we use two reference models for the relationship between concentration parameter, $c$, and halo mass, $M_{200}$. In this appendix we fitting formulae to reproduce the $c-M$ relationships.

The first relationship from {\it Prada et al.} is given by~\cite{Prada2012}
\begin{align}
    c_\textrm{prada}(M,z) &= B_0(x)\mathcal{C}(\sigma'),\\
    \sigma' &= B_1(x)\sigma(M,z),\\
    \mathcal{C}(\sigma') &= A\bigg[\bigg(\frac{\sigma'}{b}\bigg)^c + 1\bigg]\exp\bigg(\frac{d}{\sigma'}\bigg),\\
    B_0(x) &= \frac{c_\textrm{min}(x)}{c_\textrm{min}(1.393)},\\
    B_1(x) &= \frac{\sigma_\textrm{min}^{-1}(x)}{\sigma_\textrm{min}^{-1}(1.393)},\\
    c_\textrm{min}(x) &= c_0 + (c_1 - c_0)\bigg[\frac{1}{\pi}\arctan[\alpha(x-x_0)] + \frac{1}{2}\bigg],\\
    \sigma^{-1}_\textrm{min}(x) &= \sigma^{-1}_0 + (\sigma^{-1}_1 - \sigma^{-1}_0)\bigg[\frac{1}{\pi}\arctan[\beta(x-x_1)] + \frac{1}{2}\bigg], \\
    x &= \frac{1}{1+z} \bigg(\frac{\Omega_{\lambda,0}}{\Omega_{m,0}}\bigg)^{1/3}, \label{eq:xDef}
\end{align}
where the fitting constants were found to be
\begin{align}
    A &= 2.881,& b &= 1.257,& c&=1.022,& d&=0.060,\\
    c_0 &= 3.681,& c_1 &= 5.033,& \alpha &= 6.948,& x_0 &= 0.424,\\
    \sigma^{-1}_0 &= 1.047,& \sigma^{-1}_1 &= 1.646,& \beta &= 7.386,& x_1 &= 0.526.
\end{align}

The second $c$-$M$ relationship we use is the more conservative relationship from {\it Angel et al.} They found a good fit to be~\cite{Angel:2015ilq}
\begin{equation}
    c_\textrm{angel} = 2.6 \bigg(\frac{M_{200}}{10^{10} M_\odot h^{-1}}\bigg)^{-0.03} \bigg(\frac{1+z}{10}\bigg)^{-0.11}
\end{equation}
As noted in the main text, this is the fit {\it Angel et al.} find for halos with an NFW profile. They also present a fit for Einasto profiles, but we use the NFW fit to remain consistent in methodology when comparing to the {\it Prada et al.} fit.

\section{Halo Mass Function} \label{app:hmf}
The density of dark matter halos per comoving volume is estimated by finding fitting functions based on cosmological simulations. Here we follow the approach of {\it Watson et al.}~\cite{2013MNRAS.433.1230W} such that the halo mass function can be expressed as
\begin{equation}
    \frac{dn_\mathrm{halo}}{dM_{200}} = \frac{\rho_m}{M_{200}} f(\sigma) \frac{d\ln(\sigma^{-1})}{dM_{200}} \,,
\end{equation}
where $\rho_m$ is the average matter density, $\sigma^2$ is the variance of the linear density field, and $f(\sigma)$ halo multiplicity fitting function given by
\begin{equation}
f(\sigma) = A \left[\left(\frac{\beta}{\sigma} \right)^\alpha +1 \right] e^{-\gamma/\sigma^2}\,,
\end{equation}
with $A=0.282$, $\alpha=2.163$, $\beta=1.406$, and $\gamma=1.210$. For the linear density field variance we use the fit from {\it Prada et al.}~\cite{Prada2012}
\begin{equation}
    \sigma(M,z) = D(z) \frac{16.9 M_\textrm{log12}^{0.41}}{ 1 + 1.102M_\textrm{log12}^{0.20} +6.22M_\textrm{log12}^{0.333} } \,,
\end{equation}
with $M_\textrm{log12}$ defined in \eq \eqref{eq:Mlog12}. The redshift dependence is given by
\begin{equation}
    D(z) = \frac{5}{2} \bigg(\frac{\Omega_{m,0}}{\Omega_{\Lambda,0}}\bigg)^{1/3} \frac{\sqrt{1+x^3}}{x^{3/2}}\int_0^x \bigg(\frac{x^\prime }{1+x^{\prime 3}}\bigg)^{3/2}dx^\prime \,,
\end{equation}
with $x$ defined in \eq \eqref{eq:xDef}.

\bibliography{references.bib}
\end{document}